\documentclass[10pt]{article}
\usepackage{amsmath}
\usepackage{mathabx}
\usepackage{graphicx}
\usepackage{color}
\usepackage{orcidlink}
\usepackage{hyperref}
\hypersetup{colorlinks=true, linkcolor=blue, citecolor=blue, urlcolor=blue}
\usepackage{subcaption}
\RequirePackage[numbers,sort&compress]{natbib}
\begin{document}
\baselineskip=16pt

\begin{center}
\Large \textbf{Magnetized Letelier black hole in AdS spacetime }
\end{center}

\vspace{0.2cm}

\begin{center}

{\bf Ahmad Al-Badawi}\orcidlink{0000-0002-3127-3453}\\
Department of Physics, Al-Hussein Bin Talal University, 71111,
Ma'an, Jordan. \\
e-mail: ahmadbadawi@ahu.edu.jo\\

{\bf Faizuddin Ahmed}\orcidlink{0000-0003-2196-9622}\\Department of Physics, Royal Global University, Guwahati, 781035, Assam, India\\
e-mail: faizuddinahmed15@gmail.com\\

{\bf \.{I}zzet Sakall{\i}}\orcidlink{0000-0001-7827-9476}\footnote{Corresponding author}\\
Physics Department, Eastern Mediterranean University, Famagusta 99628, North Cyprus via Mersin 10, Turkey\\
e-mail: izzet.sakalli@emu.edu.tr

{\bf Sanjar Shaymatov}\orcidlink{0000-0002-5229-7657}
\\
{Institute of Fundamental and Applied Research, National Research University TIIAME, Kori Niyoziy 39, Tashkent 100000, Uzbekistan}\\{Institute for Theoretical Physics and Cosmology, Zhejiang University of Technology, Hangzhou 310023, China}\\
e-mail: {sanjar@astrin.uz}

\end{center}

\vspace{0.3cm}

\begin{abstract}
In this study, we investigate the motion of charged, neutral, and light-like particles in a magnetized black hole solution surrounded by a cloud of strings in an anti-de Sitter (AdS) background. This spacetime admits several well-known solutions as special cases, including the Letelier-AdS black hole, the Melvin spacetime, and the Schwarzschild-AdS black hole. We demonstrate that key parameters characterizing the geometry-such as the cloud of strings parameter, the magnetic field strength, and the AdS radius-significantly affect the trajectories of these particles. Our analysis shows that increasing the cloud of strings parameter weakens the gravitational influence, while the magnetic field introduces additional attractive components that can destabilize particle orbits. We examine the photon sphere and calculate the black hole shadow, showing that the shadow radius decreases with increasing magnetic field parameter but increases with the cloud of strings parameter. These findings provide potential observables for distinguishing between different black hole models in realistic astrophysical environments.
\end{abstract}

\section{Introduction} \label{isec1}
{\color{black}Black holes (BHs) have such intense gravitational fields that even light cannot escape their event horizons. The theoretical understanding of BHs has evolved significantly since the pioneering work of Schwarzschild, with various modifications and extensions to incorporate additional physical effects such as electric charge, rotation, and cosmological constants. In recent decades, the presence of external magnetic fields and cosmic strings around BHs has garnered considerable attention, as these elements introduce unique physical phenomena and modifications to BH properties. Within the framework of general relativity (GR), the gravitational collapse of a massive star in its final evolutionary stage is considered the primary mechanism for the formation of astrophysical BHs. Their unique gravitational, thermodynamical, and astronomical properties have made them objects of intense scientific interest. Among these properties, the behavior of electromagnetic fields, particularly magnetic fields, in the vicinity of BHs continues to be a compelling area of research. Standard GR models predict that any magnetic field associated with the collapsing massive object should decay over time as $t^{-1}$ \cite{Ginzburg1964,Anderson70}. This implies that a BH would not retain an intrinsic magnetic field. Nevertheless, magnetic fields can be present due to external sources, most notably from accretion disks enveloping rotating BHs~\cite{Wald74} or from the influence of companion neutron stars~\cite{Ginzburg1964}. In many astrophysical contexts, such a magnetic field can be approximated as a test field ($B\ll B_{max}$), meaning its energy density is low enough that it does not significantly alter the background spacetime geometry \cite[see, e.g.][]{Frolov10}.

The study of magnetized BHs has become increasingly important in astrophysics and theoretical physics due to the abundant evidence of strong magnetic fields surrounding astrophysical BHs, particularly in active galactic nuclei \cite{izz01,izz02,izz02x}. These magnetic fields significantly influence the dynamics of surrounding matter, affect accretion processes, and contribute to the formation of relativistic jets \cite{izz03,izz04,izz04x}. This is because even a relatively small magnetic field $B$ can profoundly impact the dynamics of charged particles, a consequence of the large Lorentz force they experience  \cite{Frolov11,Frolov12,Shaymatov15,DeLaurentis18,Shaymatov21,Shaymatov20}. Consequently, the magnetic field is gaining significance as a background field for investigating the background geometry in the vicinity of BHs. The theoretical framework for magnetized BHs was initially developed by Ernst \cite{mbhh1} and Melvin \cite{mbhh2}, who established what is now known as the Ernst-Melvin solution \cite{izz04xx}. This class of solutions describes static and spherically symmetric BHs within the Melvin magnetic universe, where the gravitational contribution of the magnetic field is considered. 
These are commonly referred to as magnetized BHs. The magnetized Reissner-Nordstr\"{o}m BH represents a prominent example of this type \cite{Gibbons13}. More recently, an alternative approach to modeling magnetized BHs was developed, specifically by incorporating global charge \cite{Gibbons14,Astorino16}. Consequently, considerable research has been devoted to understanding the diverse properties of these magnetized BH systems \cite{Konoplya,Shaymatov21rn,Shaymatov23,Shaymatov22,Shaymatov22b}. Concurrently, the exploration of BHs surrounded by cloud of strings (CS) has emerged as a significant area of research. Cosmic strings are topological defects that might have formed during phase transitions in the early universe \cite{izz05,izz06,izz06x,izz06xx,izz06xxx,izz06xxxx}. These one-dimensional objects, characterized by high energy density, create distinctive gravitational effects in their vicinity. The groundbreaking work of Letelier \cite{mbh1,mbh2} provided a mathematical framework for BHs surrounded by a CS, referred to as the Letelier solution. This solution extends the Schwarzschild metric to include the gravitational influence of a string cloud, characterized by a parameter $\alpha$ that represents the density of the cosmic strings \cite{izz07x}.

Another important development in BH physics has been the incorporation of a negative cosmological constant, leading to AdS spacetime. The AdS BHs have garnered significant interest due to their role in the AdS/CFT correspondence \cite{izz07,izz08,izz08x}, which establishes a remarkable connection between gravitational theories in AdS spacetime and conformal field theories on the boundary of this spacetime. The Schwarzschild-AdS BH serves as a fundamental solution in this context, exhibiting distinct thermodynamic properties and stability characteristics compared to its asymptotically flat counterpart \cite{izz09,izz10}. Despite the extensive research on magnetized BHs, CS, and AdS spacetime individually, their combined effects have received limited attention. The intricate interplay between these elements creates a rich physical scenario that warrants detailed investigation. The magnetized Letelier BH in AdS spacetime represents a comprehensive solution that incorporates external magnetic fields, CS, and a negative cosmological constant simultaneously. This solution reduces to several well-known cases under specific parameter choices: the Letelier-AdS spacetime when the magnetic field parameter $B_0=0$, the Ernst spacetime when the cosmic string parameter $\alpha=0$, the Schwarzschild-AdS spacetime when both $B_0=\alpha=0$, and the Melvin spacetime when $M=\alpha=0$ and the AdS radius $\ell_p \rightarrow \infty$ \cite{mbhh1,mbhh2}.

The study of particle dynamics around BHs provides valuable insights into their gravitational effects and physical properties. The trajectories of test particles-whether charged, neutral, or massless-serve as probes of the spacetime geometry, showing how different parameters influence the gravitational interaction \cite{izz11,izz12}. In the context of magnetized Letelier BH in AdS spacetime, the motion of charged particles is particularly interesting due to the combined influence of the gravitational field, the external magnetic field, the CS, and the negative cosmological constant \cite{izz13,izz14}. Similarly, the investigation of neutral particles and photon paths around these BHs illuminates their distinctive gravitational characteristics and optical properties. The optical properties of BHs, including the photon sphere and shadow, have gained prominence in recent years, especially following the remarkable achievement of the Event Horizon Telescope (EHT) in capturing the first image of a BH shadow \cite{izz15,izz16}. The photon sphere, a region where photons can orbit the BH in unstable circular orbits, determines the boundary of the BH shadow as observed from a distance. The characteristics of the photon sphere and shadow provide observational signatures that depend on the specific properties of the BH spacetime, making them valuable tools for testing different BH models \cite{izz17,izz18}.

Recent advances in numerical techniques and observational capabilities have further motivated the study of various BH solutions and their observational implications. The stability of particle orbits, characterized by Lyapunov exponents, and the periods of circular orbits provide additional metrics for understanding the dynamics around BHs \cite{VC,SF}. In particular, the presence of magnetic fields and cosmic strings introduces distinctive effects on these quantities, offering potential observational discriminators between different BH models \cite{izz19,izz20}. On the other hand, the AdS/CFT correspondence provides a powerful framework for exploring the connection between gravitational phenomena in the bulk and quantum field theory on the boundary \cite{izz21,izz22}, possibly uncovering new facets of the duality \cite{izz23,izz24}.

In this paper, we present a comprehensive analysis of the magnetized Letelier BH in AdS spacetime, investigating the influence of key parameters-the magnetic field strength $B_0$, the CS parameter $\alpha$, and the AdS radius $\ell_p$-on various physical properties. We explore the motion of charged, neutral, and light-like particles in this spacetime, examining how the combined effects of magnetization, cosmic strings, and the negative cosmological constant shape their trajectories. Our motivation stems from the recognition that real astrophysical BHs exist in environments where multiple effects coexist and interact. By incorporating magnetic fields, CS, and a negative cosmological constant within a single framework, we aim to develop a more realistic and comprehensive model of BH physics. The orthonormal components of the magnetic field, as measured by zero-angular-momentum observers (ZAMOs), provide valuable insights into the magnetic field configuration around the BH. These components reveal how the magnetic field structure is influenced by the presence of the CS and the AdS background, offering a more complete picture of the electromagnetic environment surrounding the BH \cite{izz25,izz26}. Understanding this magnetic field structure is crucial for interpreting observations of magnetized BHs and modeling phenomena such as jet formation and accretion processes \cite{izz27,izz28}. Our analysis reveals several noteworthy results: First, we demonstrate that the presence of a magnetic field significantly alters the effective potential experienced by charged particles, leading to substantial modifications in their trajectories. Second, we show that the CS parameter influences the size of the BH horizon and affects the stability of particle orbits. Third, we establish that the combined effect of magnetization and cosmic strings leads to distinctive characteristics in the photon sphere and BH shadow \cite{izz28x,izz28xx,izz28xxx,izz28xxxx}.

The paper is organized as follows: In Section \ref{isec2}, we introduce the magnetized Letelier BH solution in AdS spacetime, discussing its metric, electromagnetic properties, and relationship to other well-known spacetimes. Section \ref{isec3} analyzes the motion of charged particles in this background, examining the influence of various parameters on their trajectories and effective potentials. In Section \ref{isec4}, we investigate the behavior of neutral particles, focusing on their circular orbits and associated physical quantities. Section \ref{isec5} explores the optical properties of the magnetized Letelier BH in AdS spacetime, including photon trajectories, the photon sphere, and forces on photon particles. Section \ref{isec6} examines the BH shadow, presenting numerical findings and visualizations of its parameter-induced variations. Finally, in Section \ref{isec7}, we summarize our findings and discuss their implications for future directions.}

\section{Magnetized AdS BH Spacetime with CS} \label{isec2}

In this section, we introduce a static and axisymmetric BH spacetime in AdS background, incorporating the presence of a magnetic field and a cloud of strings (CS). We investigate in detail the geometric properties of this spacetime, focusing on the motion of neutral, charged, and light-like particles. Furthermore, we study BH shadow and analyze how various factors influence its size and shape.

 The following line element describes the geometry of the magnetized BH, which is a static, axially symmetric BH solution to the Einstein-Maxwell equations \cite{mbhh1,mbhh2} 
\begin{equation}
    ds^2=\Lambda^2 \, e^{-2\,u}\,g_{ij}\,dx^i\,dx^j +\frac{e^{2\,u}}{\Lambda^2}\,d\phi^2,\label{bb0}
\end{equation}
where the function $u$ and the metric  $g_{ij}$ depend only on  the remaining coordinates ($r,\theta$) $x^i$ with  $i, j=t, r,\theta$.

In order to construct the magnetized Letelier BH, we need to start with the Nambu-Goto action, which describes strings like objects \cite{mbh1,mbh2},   
\begin{equation}
    S^{CS}=\int \sqrt{-\gamma}\,\mathcal{M}\,d\lambda^0\,d\lambda^1=\int \mathcal{M}\sqrt{-\frac{1}{2}\,\Sigma^{\mu \nu}\,\Sigma_{\mu\nu}}\,d\lambda^0\,d\lambda^1,\label{ac1}
\end{equation}
where $\mathcal{M}$ is the dimensionless constant which characterizes the string, ($\lambda^0\,\lambda^1$) are the time
like and spacelike coordinate parameters, respectively \cite{mbh3}. $\gamma$  is the determinant of the induced metric of the strings world sheet given by $\gamma=g^{\mu\nu}\frac{ \partial x^\mu}{\partial \lambda^a}\frac{ \partial x^\nu}{\partial \lambda^b}$.  $\Sigma_{\mu\nu}=\epsilon^{ab}\frac{ \partial x^\mu}{\partial \lambda^a}\frac{ \partial x^\nu}{\partial \lambda^b}$ is bivector related to string world sheet, where $\epsilon^{ ab}$ is the second rank Levi-Civita tensor which takes the non-zero values as $\epsilon^{ 01} = -\epsilon^{ 10} = 1$.\\ The equations of motion can be obtained by varying the action (\ref{ac1}) with respect to the metric, $g^{\mu\nu}$, and the  magnetic potential, $A_\phi=\frac{B_0\,e^{2u}}{\Lambda}$ namely: 
 \begin{equation}
     G_{\mu\nu}-\frac{3}{\ell_p^2}\,g_{\mu \nu}=8 \pi\, \left( \,T_{\mu\nu}^{CS}+\,T_{\mu\nu}^{em}\right), 
 \end{equation}
where $G_{\mu\nu}$ is the Einstein tensor, $T_{\mu\nu}^{CS}$ and $T_{\mu\nu}^{em}$ are energy-momentum tensor  associated with CS and the electromagnetic field sources, respectively. The energy-momentum tensors for the CS is given by 
 \begin{equation}
   T_{\mu\nu}^{CS}=2 \frac{\partial}{\partial g_{\mu \nu}}\mathcal{M}\sqrt{-\frac{1}{2}\Sigma^{\mu \nu}\,\Sigma_{\mu\nu}} =\frac{\rho \,\Sigma_{\alpha\nu}\, \,\Sigma_{\mu}^\alpha }{\sqrt{-\gamma}}, 
 \end{equation}
where $\rho$ is the proper density of the CS. The energy-momentum tensors for the electromagnetic field is given by
\begin{equation}
   T_{\mu \nu }^{em}=\frac{1}{4 \pi} \left( F_{\mu \rho }F_{\nu}^{\rho }-\frac{1}{4}g_{\mu \nu }F^2 \right). 
\end{equation}
We can obtain the non-vanishing components of the energy-momentum tensor of CS by applying conservation of law, $\nabla_\mu \, T_{\mu\nu}^{CS}=0 $, thus 
\begin{equation}
 T_{t}^{CSt}=T_{r}^{CSr}=-\frac{\alpha}{r^2},
\end{equation}  
where the constant $\alpha$ represent the CS parameter.

When applying the magnetized technique \cite{mbh4} to the Letelier solution, one should note that: 
\begin{equation}
    e^{2u}=r^2\,\sin^2\theta, \hspace{1cm} e^{-2u}\,g_{ij}\,dx^i\,dx^j=-\mathcal{F}(r)\,dt^2+\frac{dr^2}{\mathcal{F}(r)}+r^2\,d\theta^2
\end{equation}
In this case, one could use the solution generating techniques developed in Ref. \cite{mbh4} to obtain the following new exact magnetized Letelier BH solution:
\begin{equation}
    ds^2=\Lambda^2 \left[-\mathcal{F}(r)\,dt^2+\frac{dr^2}{\mathcal{F}(r)}+r^2\,d\theta^2 \right]+\frac{r^2\,\sin^2\theta}{\Lambda^2}\,d\phi^2,\label{bb1}
\end{equation}
where 
\begin{equation}
  \Lambda=\Lambda(r)=1+B^2_0 \,r^2\, \sin^2\theta, \hspace{1cm}  \mathcal{F}(r)=1-\alpha-\frac{2\,M}{r} + \frac{r^2}{\ell^2_p}.\label{bb2}
\end{equation}
Here, the parameter $B_0$  is a constant determines the strength of the external magnetic field.  Our solution extends magnetized Letelier BH in AdS spacetime ($\alpha \ne 0$) to include CS. As boundary conditions, our metric (\ref{bb1}) reduces to some well-known solutions under several limits. A few are as follows:  
\begin{eqnarray*}
&&\,\,\mbox{when}\,\, B_0=0, \,\, \rightarrow\text{Letelier AdS spacetime}\,\cite{mbh1},  
 \\
&&\,\,\mbox{when}\,\, \alpha=0, \,\,\rightarrow\text{ Ernst spacetime} \,\cite{mbhh1}\\
&&\,\,\mbox{when}\,\, B_0 =0=\alpha, \,\, \rightarrow\text{Sch. AdS spacetime}\,\cite{TVF}\\
&&\,\,\mbox{when}\,\,\ell_p \rightarrow \infty,\,\, M=0=\alpha, \,\, \rightarrow  \text{Melvin spacetime} \, \cite{mbhh2}. \label{bb16c}
\end{eqnarray*}

\begin{figure*}[ht!]
\includegraphics[width=0.3\textwidth]{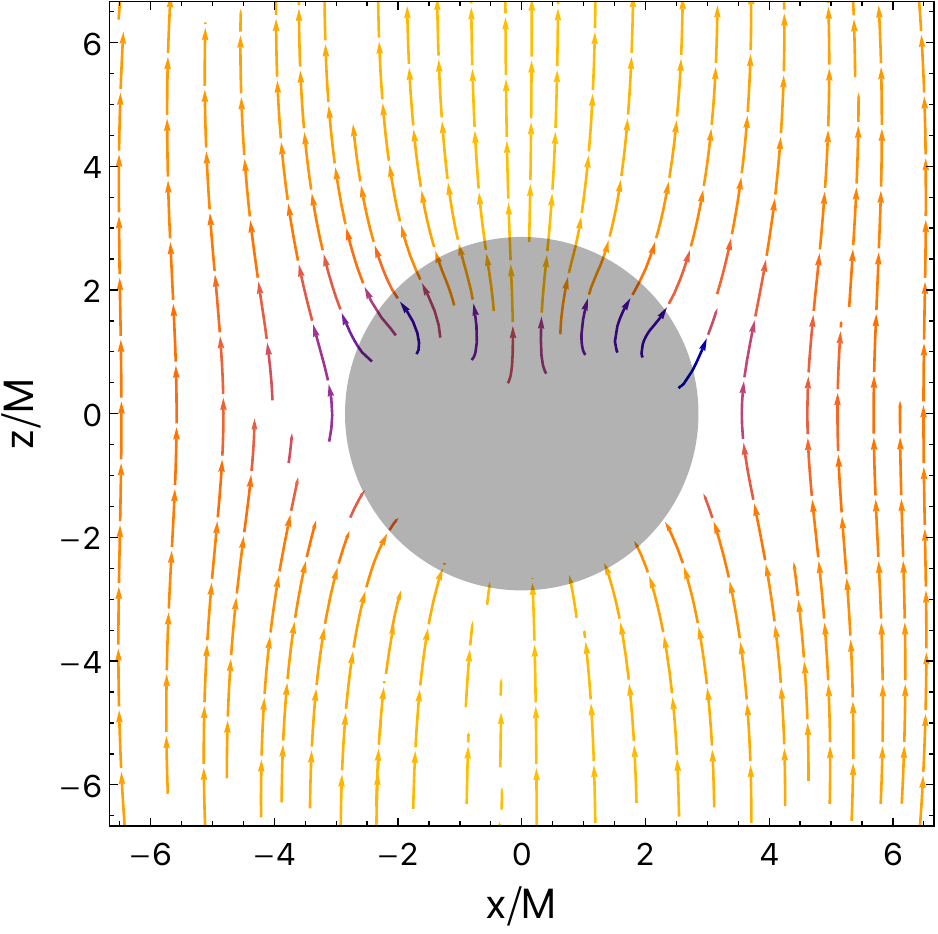}
\includegraphics[width=0.3\textwidth]{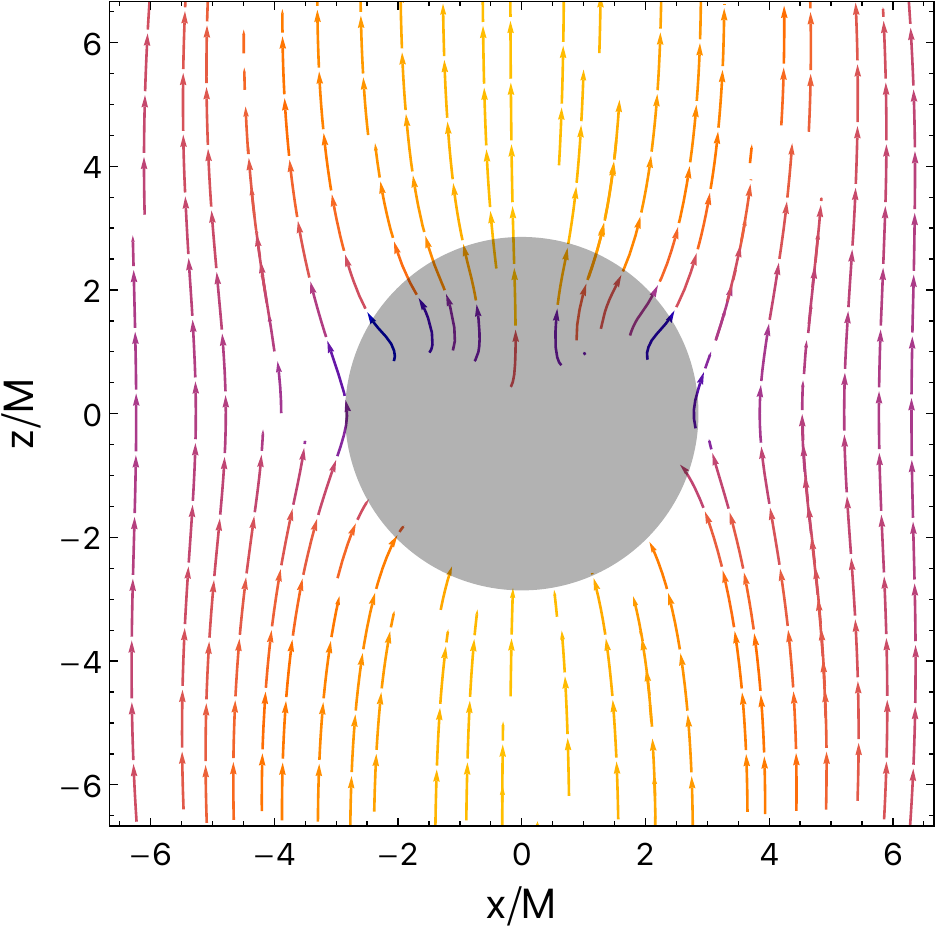}
\includegraphics[width=0.3\textwidth]{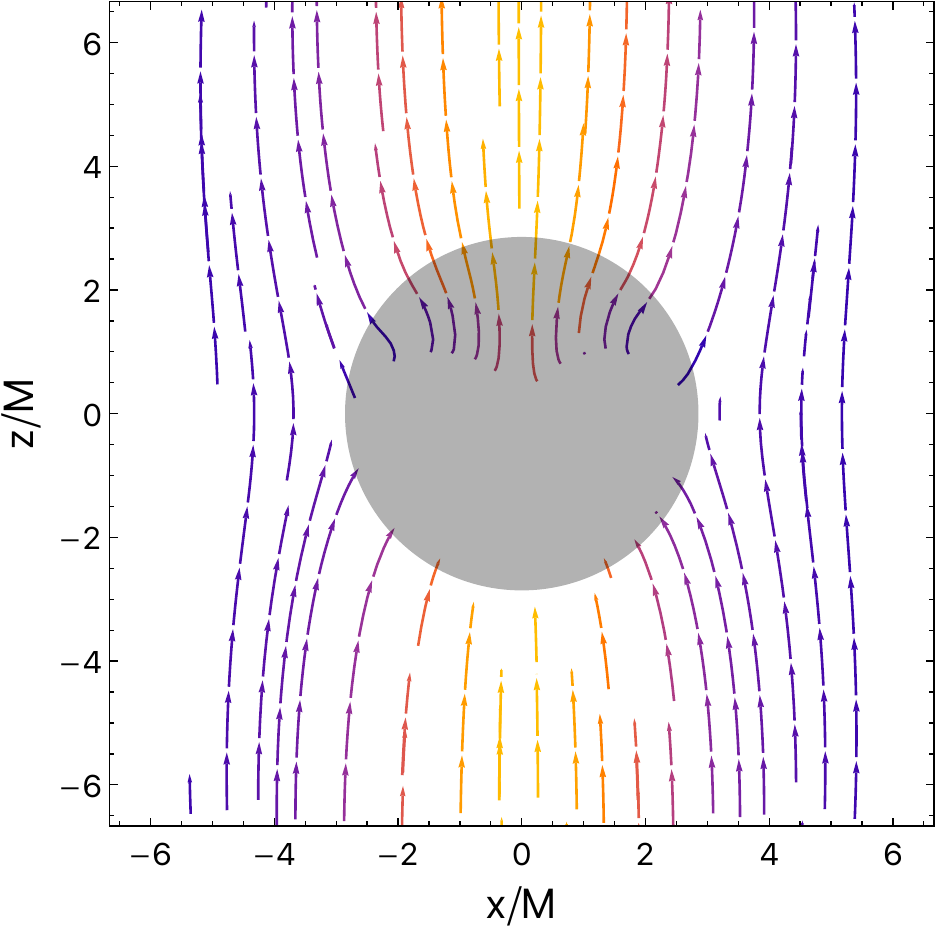} 
\caption{\label{Fig:MF} Plot shows the configuration of magnetic field lines in the environment of the magnetized Letelier BH in AdS spacetime for various combinations of magnetic field parameter $B=0.05,\,0.1,\,0.5$ while keeping $\alpha=0.3$ and $\ell_p=10$ fixed, corresponding to panels from left to right. Note that the gray-shaded area refers to the BH horizon. Note that the magnetic field parameter has been considered a dimensionless quantity $B_0\to B_0M$ by setting $G = c = 1$).}
\end{figure*}

It is worth noting that the magnetized Letelier BH solution is neither asymptotically flat nor spherically symmetric. Interestingly, the event horizon is not affected by the magnetic field $B_0$, similarly to what was obtained for the Schwarzschild AdS and the Schwarzschild spacetime BH $\alpha=\ell_p=0$. Here, we can define the electromagnetic field around the magnetized Letelier BH as follows:
\begin{align}\label{eq:vec-pot}
A_{\mu}dx^{\mu}=\frac{B_0 r^2 \sin^2\theta}{2\Lambda}d\phi\, .
\end{align}
The assumed axial magnetic field $B_0$ breaks the spacetime's spherical symmetry, resulting in an axisymmetric geometry. The orthonormal components of the magnetic field, as measured by zero-angular-momentum observers (ZAMOs) with four-velocity components $\left(u_{ZAMO}\right)^{\mu}=\{(\Lambda^2\mathcal{F}(r))^{-1/2},0,0,0\}$ and $\left(u_{ZAMO}\right)_{\mu}=\{(\Lambda^2\mathcal{F}(r))^{1/2},0,0,0\}$, are given by:
\begin{eqnarray}
\label{b1}  B^{\hat r}
&&=-\frac{B_0}{\Lambda}\left(1-\frac{B^2r^2\sin^2\theta}{\Lambda}
\right)\cos\theta \, , \\
\label{b2}  B^{\hat\theta} &&
=\frac{B_0\mathcal{F}(r)^{1/2}}{\Lambda}\left(1-\frac{B^2r^2\sin^2\theta}{\Lambda}
\right)\sin\theta.
\end{eqnarray}
As can be seen from Eqs.~(\ref{b1}) and (\ref{b2}), the magnetic field components are determined by the parameter that defines the external magnetic field. In the limit of \hbox{$M/r\rightarrow 0$}, $\alpha\to 0$, $\ell_p \rightarrow \infty$ and
\hbox{$\Lambda\rightarrow 1$}, the solutions will be reduced to the
flat spacetime solutions 
\begin{eqnarray}
 B^{\hat r} =-B_0\cos\theta,  ~~~ B^{\hat\theta}=B_0\sin\theta \, .
\end{eqnarray}
This coincides, as expected, with a homogeneous magnetic field in Newtonian spacetime. Fig.~\ref{Fig:MF} depicts the magnetic field line configuration in the vicinity of the magnetized Letelier BH in AdS spacetime.

\begin{figure}
\centering
\includegraphics[scale=0.9]{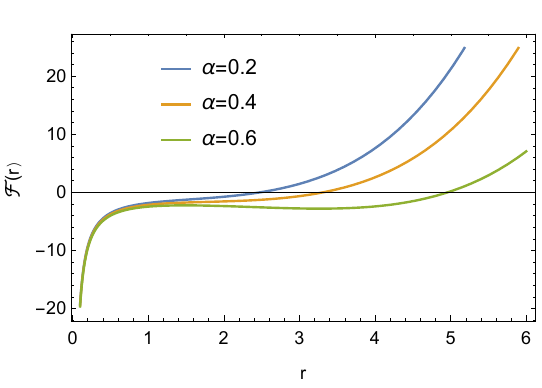}\includegraphics[scale=0.9]{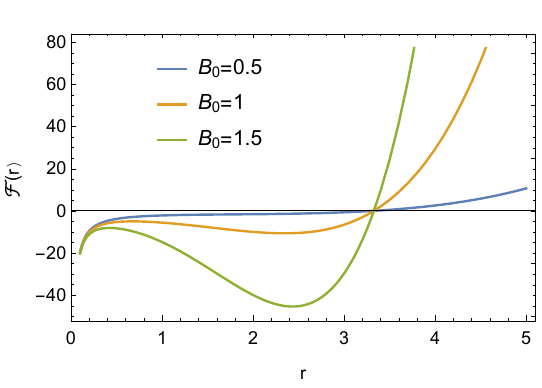}
\caption{Metric function $\mathcal{F}(r)$ for varying values of $\alpha$ by setting $B_0=0.5$ (left) and for varying values of $B_0$ by setting $\alpha=0.4$ (right). Here, $M=1$ and $\ell_p=100$.}
\label{figa1}
\end{figure}

To shed light on the metric function as well as the  horizon  of magnetized BH, we generate  Fig.~\ref{figa1}. The Figure  depicts how parameters $\alpha$ and $B_0$ affect the metric  function $\mathcal{F}(r)$ for various values. The figure demonstrates that magnetized Letelier BH in AdS spacetime has only one unique  horizon. As the CS parameter increases the horizon grows. This plot clearly shows how CS  parameter plays a significant role in BH horizon existence. On the other hand, the intensity of magnetic field has no effect on BH horizon.     

\section{Motion of charged particles in magnetized Letelier AdS spacetime} \label{isec3}

In this section, we investigate the motion of a charged particle under the influence of the gravitational field generated by a magnetized BH, highlighting how various factors in the space-time geometry affect its dynamics. To analyze this behavior, we consider the motion of the charged particle confined to the equatorial plane, defined by $\theta = \frac{\pi}{2}$, and employ the Lagrangian formalism to study the system in a systematic manner.

For a charged particle with mass $\mu$ and electric charge $q$, it’s motion can be described using the following Lagrangian  density function
\begin{equation}
    \mathcal{L}=\frac{1}{2}\,g_{\mu\nu}\,\dot{x}^{\mu}\,\dot{x}^{\nu}+\epsilon\,A_{\mu}\,\dot{x}^{\mu},\label{cc1}
\end{equation}
where $\epsilon=q/\mu$ is the charge-to-mass ratio and dot represents partial derivative w. r. to an affine parameter $\lambda$, and $A_{\mu}$ is the electromagnetic four-vector potential.

Using the metric (\ref{bb1}) and potential component (\ref{eq:vec-pot}), we find the Lagrangian density function in the equatorial plane $\theta=\pi/2$ as
\begin{equation}
    \mathcal{L}=\frac{1}{2}\,\left[-\tilde{\Lambda}^2\,\mathcal{F}(r)\,\dot{t}^2+\frac{\tilde{\Lambda}^2}{\mathcal{F}(r)}\,\dot{r}^2+\frac{r^2}{\tilde{\Lambda}^2}\,\dot{\phi}^2\right]+\epsilon\,\tilde{A}_{\phi}\,\dot{\phi}.\label{cc2}
\end{equation}

Since the Lagrangian density function is independent of $t$ and $\phi$, there are two constants of motions $\mathrm{E}$ and $\mathrm{L}$. There are related with the geodesics equations of $t$ and $\phi$ given by
\begin{equation}
    \dot{t}=\frac{\mathrm{E}}{\tilde{\Lambda}^2\,\mathcal{F}},\quad\quad\quad \dot{\phi}=\frac{\tilde{\Lambda}^2\,(\mathrm{L}-\epsilon\,\tilde{A}_{\phi})}{r^2},\quad\quad\quad \tilde{A}_{\phi}=\frac{r^2\,B_0}{1+B^2_0\,r^2},\quad\quad\quad \tilde{\Lambda}=1+B^2_0\,r^2.\label{cc3}
\end{equation}

Substituting $\dot{t}$ and $\dot{\phi}$ into the Eq. (\ref{cc2}), and $2\,\mathcal{L}=-1$ for time-like particles and after some straightforward simplification yields:
\begin{equation}
    \tilde{\Lambda}^4\,\dot{r}^2+V_\text{eff}(r)=\mathrm{E}^2,\label{cc4}
\end{equation}
where $V_\text{eff}(r)$ is the effective potential given by
\begin{equation}
    V_\text{eff}(r)=(1+B^2_0\,r^2)^2\,\left(1-\alpha-\frac{2\,M}{r}+\frac{r^2}{\ell^2_p}\right)\,\left[1+\frac{(1+B^2_0\,r^2)^2}{r^2}\,\left\{\mathrm{L}^2-\frac{\epsilon^2\,B^2_0\,r^4}{(1+B^2_0\,r^2)^2}\right\}\right].\label{cc5}
\end{equation}

From expression (\ref{cc5}), it is clear that the effective potential for charged particle is influenced by several factors involved in the space-time geometry. These include the magnetic field strength $B_0$, the CS parameter $\alpha$, the AdS radius $\ell_p$. Additionally, the BH mass $M$ and the charge per unit mass $\epsilon$ also alters this effective potential.

\begin{figure}[ht!]
    \centering
    \subfloat[$B_0=0.1$]{\centering{}\includegraphics[width=0.45\linewidth]{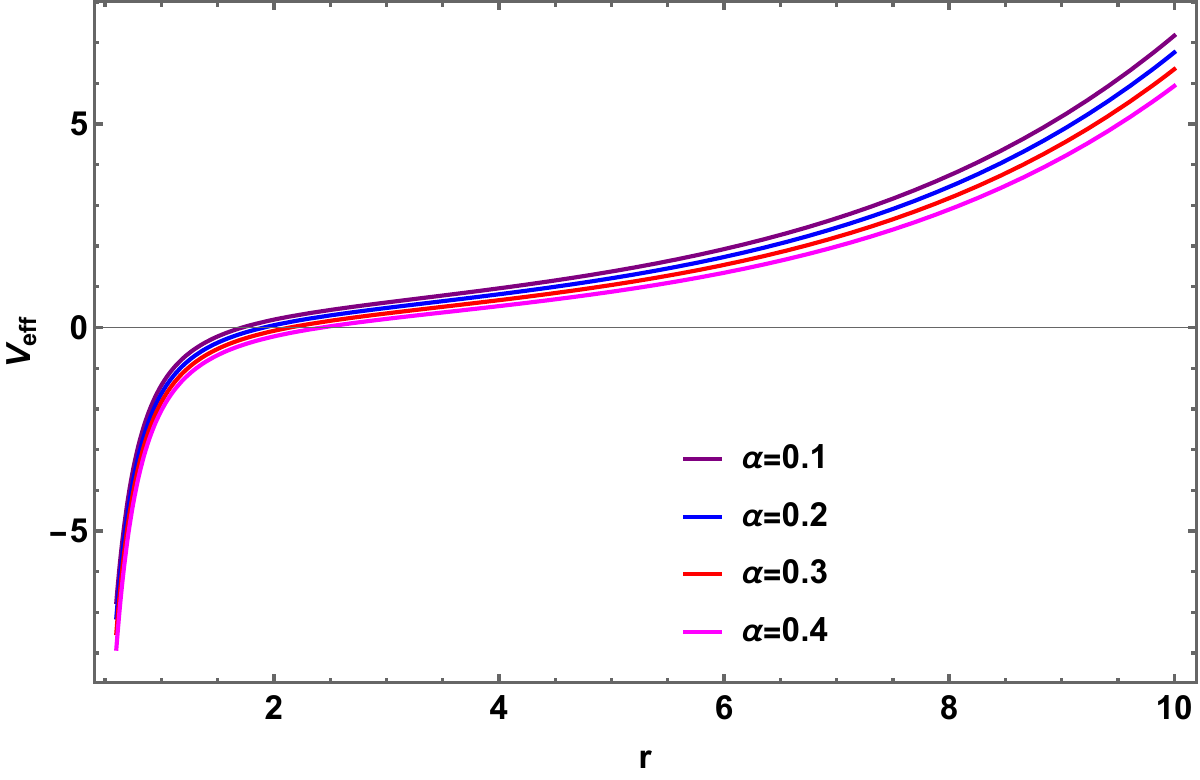}}\quad\quad
    \subfloat[$\alpha=0.1$]{\centering{}\includegraphics[width=0.45\linewidth]{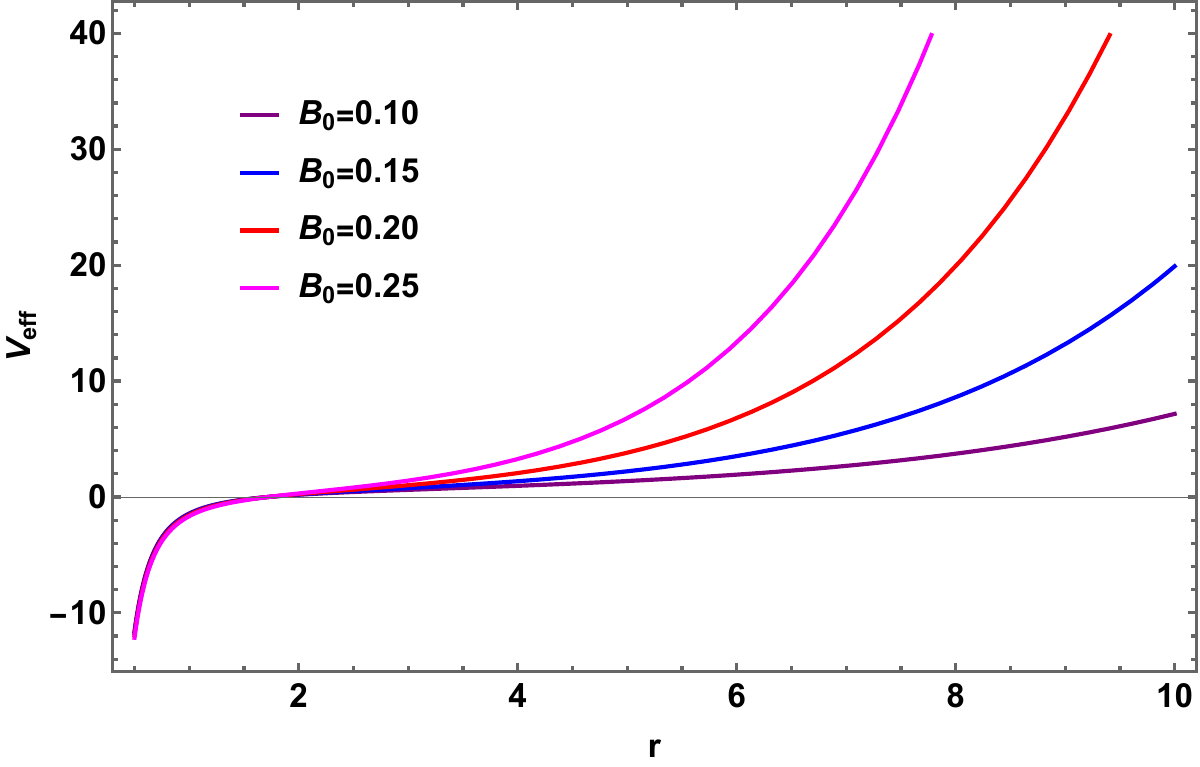}}\\
    \subfloat[]{\centering{}\includegraphics[width=0.45\linewidth]{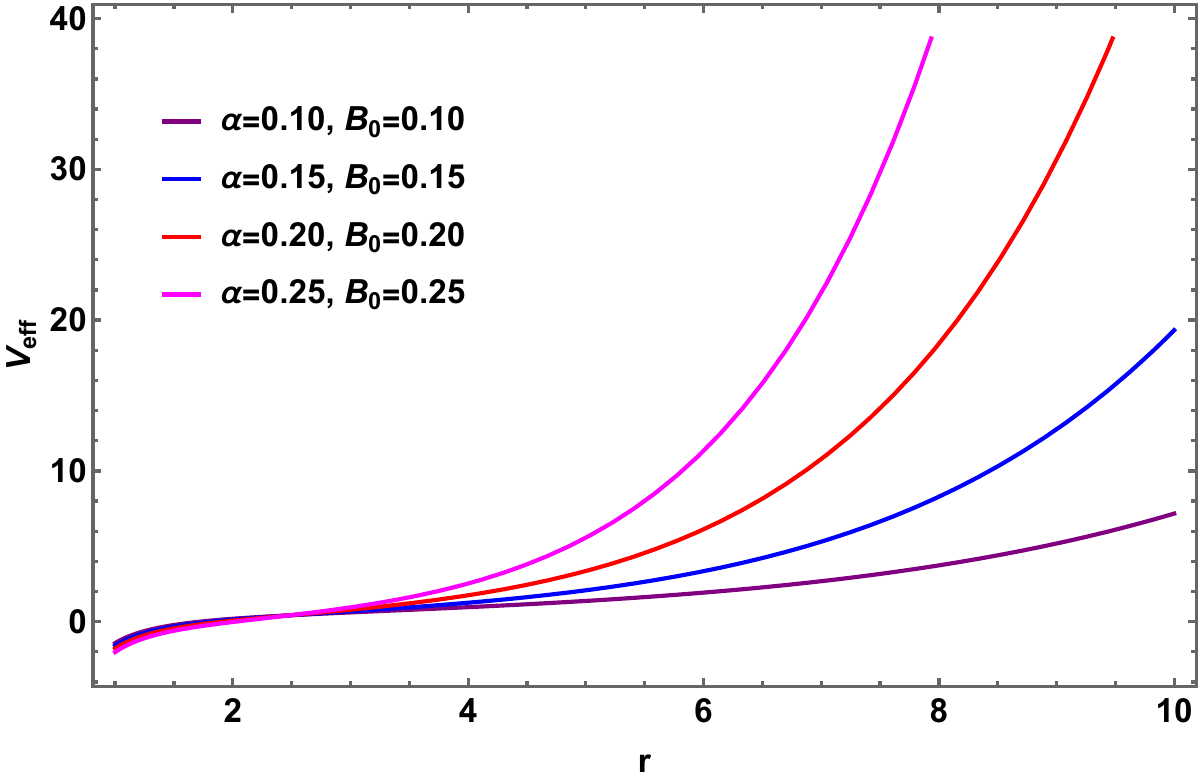}} 
    \caption{The behavior of the effective potential for charged particle by varying $\alpha$ and $B_0$. Here $M=0.8$, $\mathrm{L}=1$, $\epsilon=0.1$, and $\ell_p=10$.}
    \label{fig:potential1}
\end{figure}

\begin{figure}[ht!]
    \centering
    \includegraphics[width=0.45\linewidth]{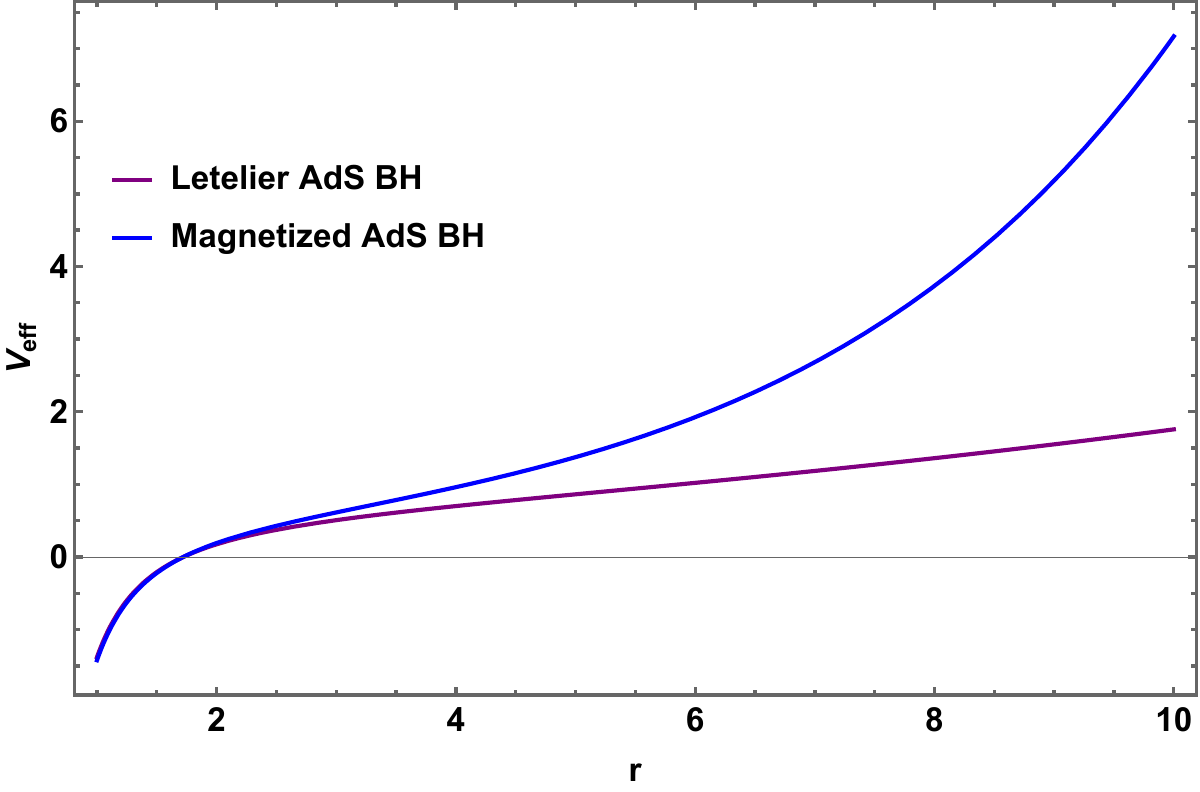}
    \caption{A comparison of the effective potential for charge particle. Here $M=0.8$, $\mathrm{L}=1$, $\epsilon=0.1$, and $\ell_p=10$. Purple color: $\alpha=0.1, B_0=0$, blue color: $B_0=0.1, \alpha=0.1$.}
    \label{fig:potential2}
\end{figure}

In the limit where $B_0 \to 0$, that is, there is no magnetic field effect on test particles, the effective potential from Eq. (\ref{cc5}) becomes
\begin{equation}
    V_\text{eff}(r)=\left(1-\alpha-\frac{2\,M}{r}+\frac{r^2}{\ell^2_p}\right)\,\left(1+\frac{\mathrm{L}^2}{r^2}\right).\label{cc6}
\end{equation}
Equation (\ref{cc6}) is the well-known effective potential expression for time-like particles in the Letelier AdS BH space-time. 

Thereby, from expressions (\ref{cc5}) and (\ref{cc6}), it becomes evident that the presence of magnetic field causes a large changes in the effective potential of the system.

\begin{figure*}[ht!]
  \includegraphics[width=0.3\textwidth]{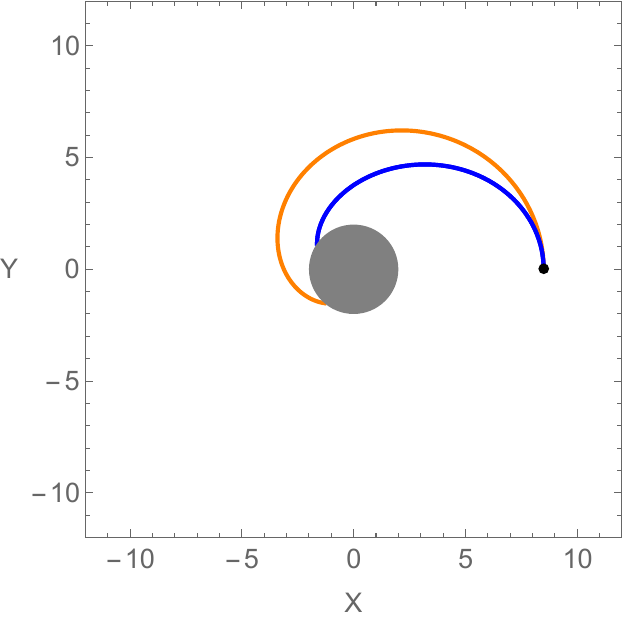}
 \includegraphics[width=0.3\textwidth]{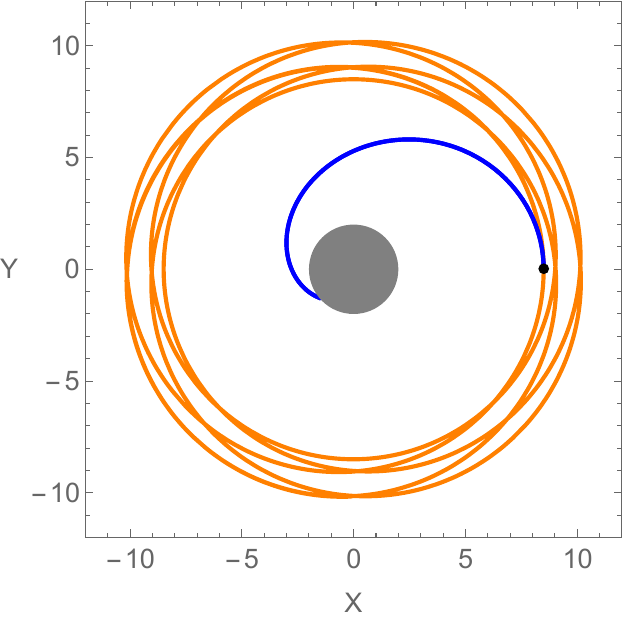}
\includegraphics[width=0.3\textwidth]{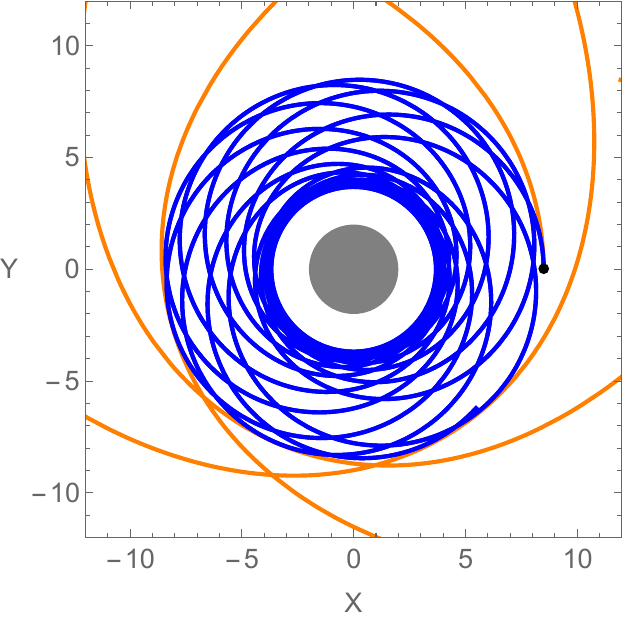}\\
\includegraphics[width=0.3\textwidth]{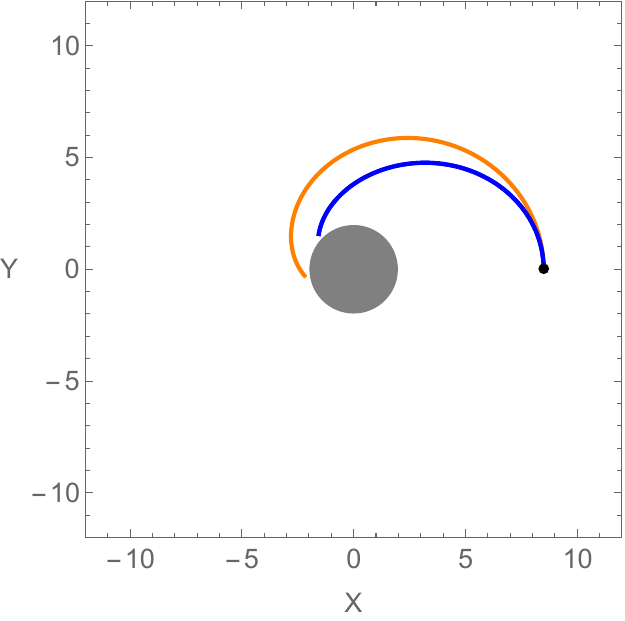}
\includegraphics[width=0.3\textwidth]{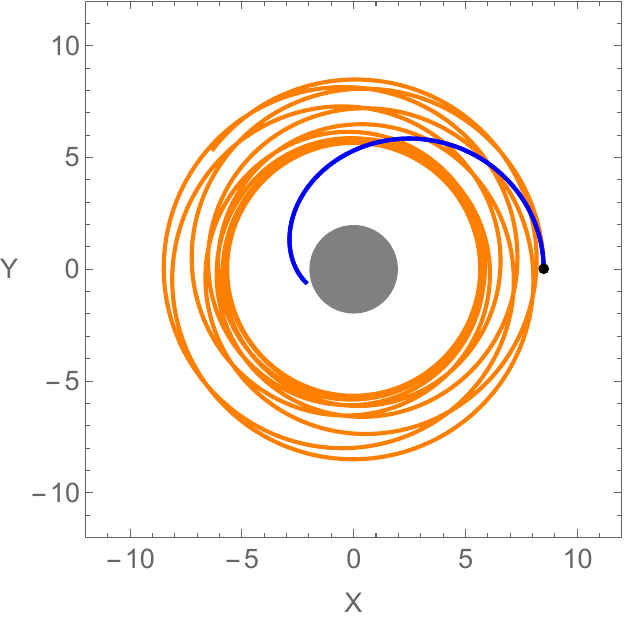}
\includegraphics[width=0.3\textwidth]{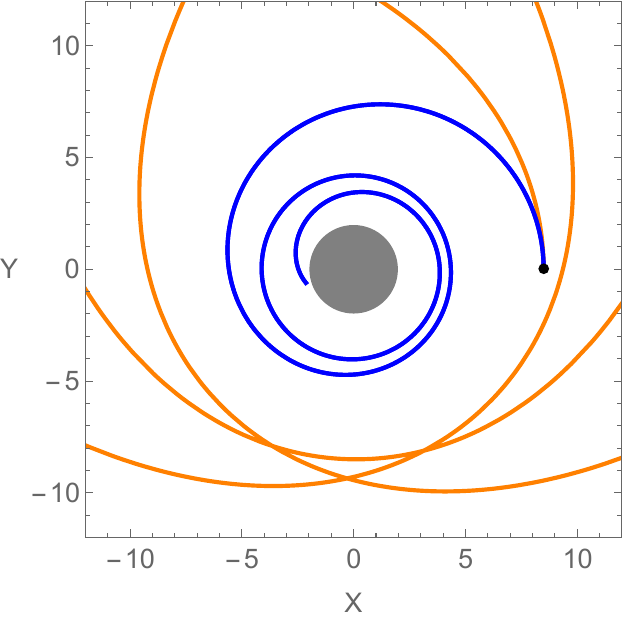} 
 \caption{\label{Fig:traj} Plot shows trajectories of test particles around the magnetized Letelier BH in AdS spacetime for magnetic field parameter $B=0.0$ (orange curve) and $B=0.05$ (blue curve) for various combinations of angular momentum $L=3,\,4,\,5$ while keeping $\ell_p=100$ fixed, corresponding to panels from left to right. Note that top/bottom row refers to $\alpha=0.0/0.1$.}
\end{figure*}
In Fig. \ref{fig:potential1}, we present the behavior of the effective potential for a charged particle under variations of the cosmic string (CS) parameter $\alpha$ and the magnetic field strength $B_0$. Panel (a) demonstrates that increasing the CS parameter from $\alpha = 0.1$ leads to a decrease in the effective potential as a function of the radial coordinate $r$. This behavior indicates that larger values of $\alpha$ effectively weaken the gravitational field influence, thereby diminishing the confining effect on the charged particle. Conversely, panels (b) and (c) show that slight increases in $B_0$, or the combined $(\alpha, B_0)$ influence result in a rise in the effective potential with increasing radial distance. This suggests that the joint contribution of the CS parameter and magnetic field amplifies the gravitational field generated by the magnetized BH, thus enhancing the potential experienced by the charged particle. In this Figure, the BH mass is fixed at $M = 0.8$, the angular momentum at $\mathrm{L} = 1$, the charge-to-mass ratio of the particle at $\epsilon = 0.1$, and the AdS radius at $\ell_p = 10$. These parameters collectively determine the behavior of the charged particles and influence whether they are captured by or escape from the gravitational influence of the BH.

In Fig. \ref{fig:potential2}, we present a comparison of the effective potential for a charged particle in the presence and absence of a magnetic field. It is observed that a slight increase in the magnetic field strength, from $B_0 = 0$ to $B_0 = 0.1$, leads to a noticeable enhancement in the variation of the effective potential experienced by the charged particle. In this Figure, the BH mass is fixed at $M = 0.8$, the angular momentum at $\mathrm{L} = 1$, the charge-to-mass ratio of the particle at $\epsilon = 0.1$, and the AdS radius at $\ell_p = 10$. 

We now turn to analyze the trajectories of test particles around a magnetized Letelier BH in AdS spacetime, focusing on motion within the equatorial plane. Fig.~\ref{Fig:traj} illustrates various particle trajectory behaviors. A deeper understanding of possible orbits around a BH is crucial. Therefore, we examine captured orbits (left), bound orbits (middle), and escape orbits (right). The middle panels of Fig.~\ref{Fig:traj} display bound orbits, arising from the balance between centrifugal and gravitational forces influenced by BH parameters. However, these orbits turn to be captured as one includes the magnetic field parameter $B$. In contrast, the right panels show no bound orbits, as the centrifugal force dominates the gravitational force when the magnetic field parameter $B$ is zero. This results in a repulsive force, allowing particles to escape the BH's pull. However, the presence of a non-zero magnetic field parameter ($B \neq 0$) introduces an attractive component, destabilizing particle orbits and leading to bounded orbits, as shown in the left and middle panels of Fig.~\ref{Fig:traj}. This analysis enhances our understanding of the magnetic field parameter in the vicinity of the magnetized Letelier BH in AdS spacetime. It should also be emphasized that from trajectories of particles, as shown in the bottom row of Fig.~\ref{Fig:traj}, the orbits are initially bounded for larger angular momentum $L$, eventually being captured due to the inclusion of the string cloud parameter $\alpha=0$. We can infer that the combined influence of the magnetic field $B$ and the string cloud parameter $\alpha$ can shift the particle orbits around the BH from bound states to captured states.

\section{Motion of neutral particles in magnetized Letelier AdS spacetime} \label{isec4}

In this section, we study the motion of neutral particles (specifically time-like) in the background of the magnetized Leterlier AdS BH space-time and analyze the outcomes.

The Hamiltonian describing the motion of a neutral particle is expressed by \cite{AB1,AB2,AB3,AB4,AB5,AB6}
\begin{equation}
    \mathrm{H}=\frac{1}{2}\,g^{\sigma\kappa}\,p_{\sigma}\,p_{\kappa}+\frac{1}{2}\,\mu^2,\label{ss1}
\end{equation}
where $\mu$ is the mass of neutral particle, $p^{\sigma}=\mu\,u^{\sigma}$ is the four-momentum, $u^{\sigma}=dx^{\sigma}/d\tau$ is the four-velocity equation, and $\tau$ is the appropriate time of the neutral particle. Also, the Hamilton equations of motion are given: 
\begin{equation}
    \frac{dx^{\sigma}}{d\lambda}\equiv \mu\,u^{\sigma}=\frac{dH}{dp_{\sigma}},\label{ss2}
\end{equation}
and
\begin{equation}
    \frac{dp_{\sigma}}{d\lambda}=-\frac{\partial H}{\partial x^{\sigma}},\label{ss3}
\end{equation}
where the affine parameter is given by $\lambda=\tau/\mu$.

Using the normalization condition $g_{\sigma\kappa}\,u^{\sigma}\,u^{\kappa}=-\epsilon$, we find
\begin{equation}
    -\Lambda^2\,\mathcal{F}(r)\,\dot{t}^2+\frac{\Lambda^2}{\mathcal{F}(r)}\,\dot{r}^2+\frac{r^2\,\sin^2 \theta}{\Lambda^2}\,\dot{\phi}^2+r^2\,\Lambda^2(r)\,\dot{\theta}^2=\epsilon,\label{ss4}
\end{equation}
where $\epsilon=0$ for null and $-1$ for time-like particle.

Since the metric components are independent of the temporal $t$ and azimuthal $\phi$ coordinates, so the particle’s four-momentum, {\it i.e.,} $p_t$ and $p_{\phi}$ are conserved along its geodesics and are given by 
\begin{eqnarray}
    &&\frac{p_t}{\mu}=-\Lambda^2\,\mathcal{F}\,\dot{t}=-\mathcal{E},\nonumber\\
    &&\frac{p_{\phi}}{\mu}=\frac{r^2\,\sin^2 \theta}{\Lambda^2}\,\dot{\phi}=\mathcal{L}_0,\label{ss5} 
\end{eqnarray}
where $\mathcal{E}=\mathrm{E}/\mu$ and $\mathcal{L}_0=\mathrm{L}/\mu$, respectively are the specific energy and angular momentum per unit mass of the neutral particle. Moreover, the conjugate momentum associated with $\theta$ coordinate is given by
\begin{equation}
    \frac{p_{\theta}}{\mu}=r^2\,\Lambda^2\,\dot{\theta}.\label{zz1}
\end{equation}

The four-velocity components $u^i$ of time-like particles, including the temporal ($u^t$), azimuthal ($u^{\phi}$) and radial ($u^r$) components, obey the following equations of motion:
\begin{eqnarray}
    &&\frac{dt}{d\tau}=\frac{\mathcal{E}}{\Lambda^2\,\mathcal{F}},\label{zz2}\\
    &&\Lambda^4\,\left(\frac{dr}{d\tau}\right)^2+\left(\Lambda^2+\frac{\Lambda^4\,\mathcal{L}^2_0}{r^2\,\sin^2 \theta}+\frac{p^2_{\theta}}{\mu^2\,r^2}\right)\,\mathcal{F}(r)=\mathcal{E}^2,\label{zz3}\\
    &&\frac{d\theta}{d\tau}=\frac{p_{\theta}}{\mu\,r^2\,\Lambda^2},\label{zz4}\\
    &&\frac{d\phi}{d\tau}=\frac{\mathcal{L}_0\,\Lambda^2}{r^2\,\sin^2 \theta},\label{zz5}
\end{eqnarray}
where the four-velocity satisfies the time-like condition $u^i\,u_i=-1$.

\subsection{Effective potential and orbital dynamics}

Since $\Lambda=\Lambda(r,\theta)$ depends on the radial and angular coordinates ($r, \theta$), we must therefore consider the motion of particles in the equatorial plane, defined by $\theta=\pi/2$ and $\frac{d\theta}{d\tau}=0$. Consequently, Eq. (\ref{zz3}) turns out to be
\begin{equation}
    \tilde{\Lambda}^4\,\dot{r}^2+U_\text{eff}(r)=\mathcal{E}^2\quad\mbox{where}\quad \tilde{\Lambda}=1+B^2_0\,r^2.\label{ss6}
\end{equation}
Here $U_\text{eff}(r)$ is the effective potential of the system in the equatorial plane and is given by
\begin{equation}
    U_\text{eff}(r)=\left(1+B^2_0\,r^2\right)^2\,\left\{1+\frac{\left(1+B^2_0\,r^2\right)^2}{r^2}\,\mathcal{L}^2_0\right\}\,\left(1-\alpha-\frac{2\,M}{r} + \frac{r^2}{\ell^2_p}\right).\label{ss7}
\end{equation}
Others geodesic paths reduces as,
\begin{eqnarray}
    &&\frac{dt}{d\tau}=\frac{\mathcal{E}}{\tilde{\Lambda}^2\,\mathcal{F}},\label{zz6}\\
    &&\frac{d\phi}{d\tau}=\frac{\mathcal{L}_0\,\tilde{\Lambda}^2}{r^2}.\label{zz7}
\end{eqnarray}

From the above expression (\ref{ss7}), it is evident that the effective potential of time-like neutral particles is influenced by mainly two key factors: the CS parameter $\alpha$ and magnetic field strength $B_0$ including the radius of curvature $\ell_p$. This effective potential $U_\text{eff}(r)$ describes the motion of neutral particles, as it illustrates their trajectories without directly solving the equations of motion. 

\begin{figure}[ht!]
    \centering
    \subfloat[$B_0=0.1$]{\centering{}\includegraphics[width=0.45\linewidth]{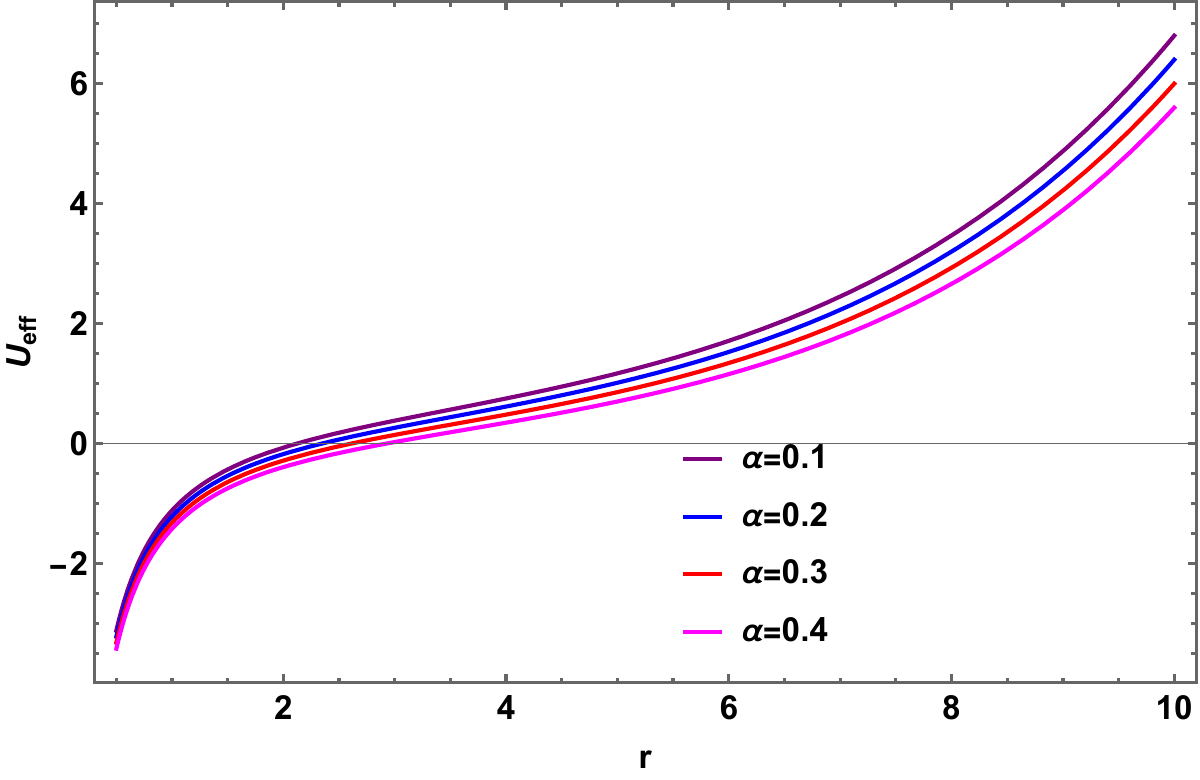}}\quad\quad
    \subfloat[$\alpha=0.1$]{\centering{}\includegraphics[width=0.45\linewidth]{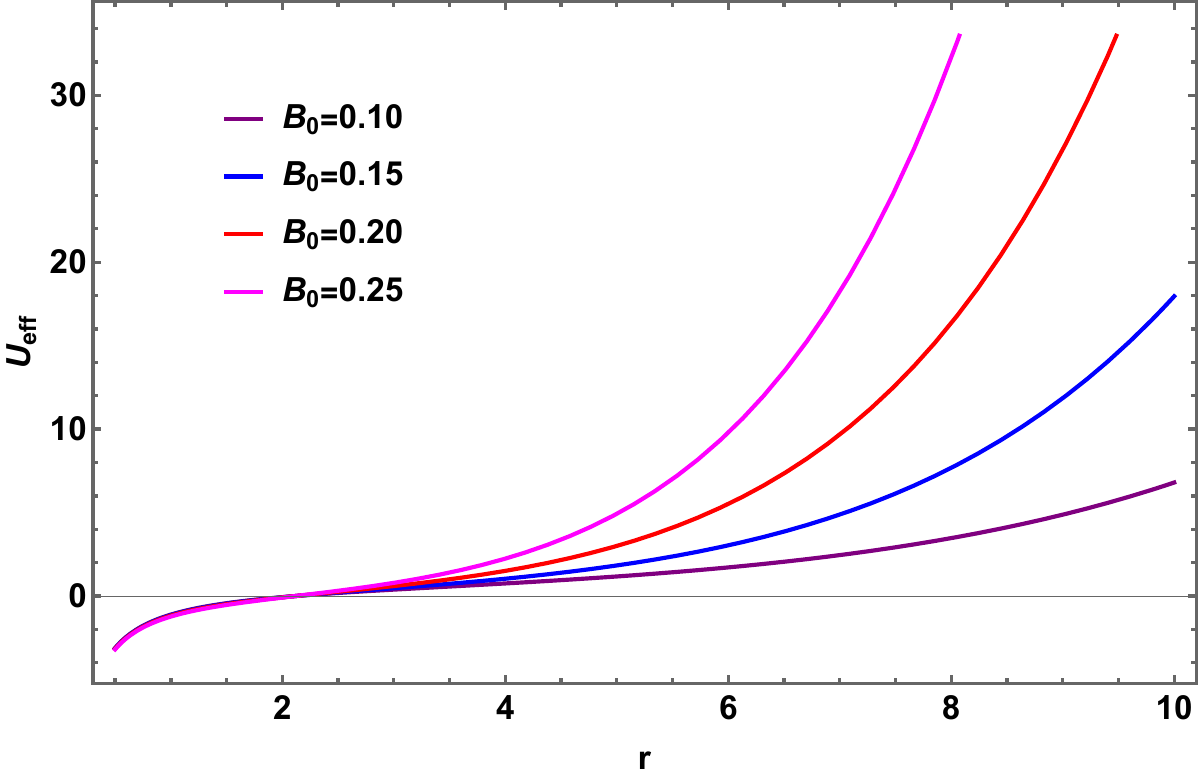}}\\
    \subfloat[]{\centering{}\includegraphics[width=0.45\linewidth]{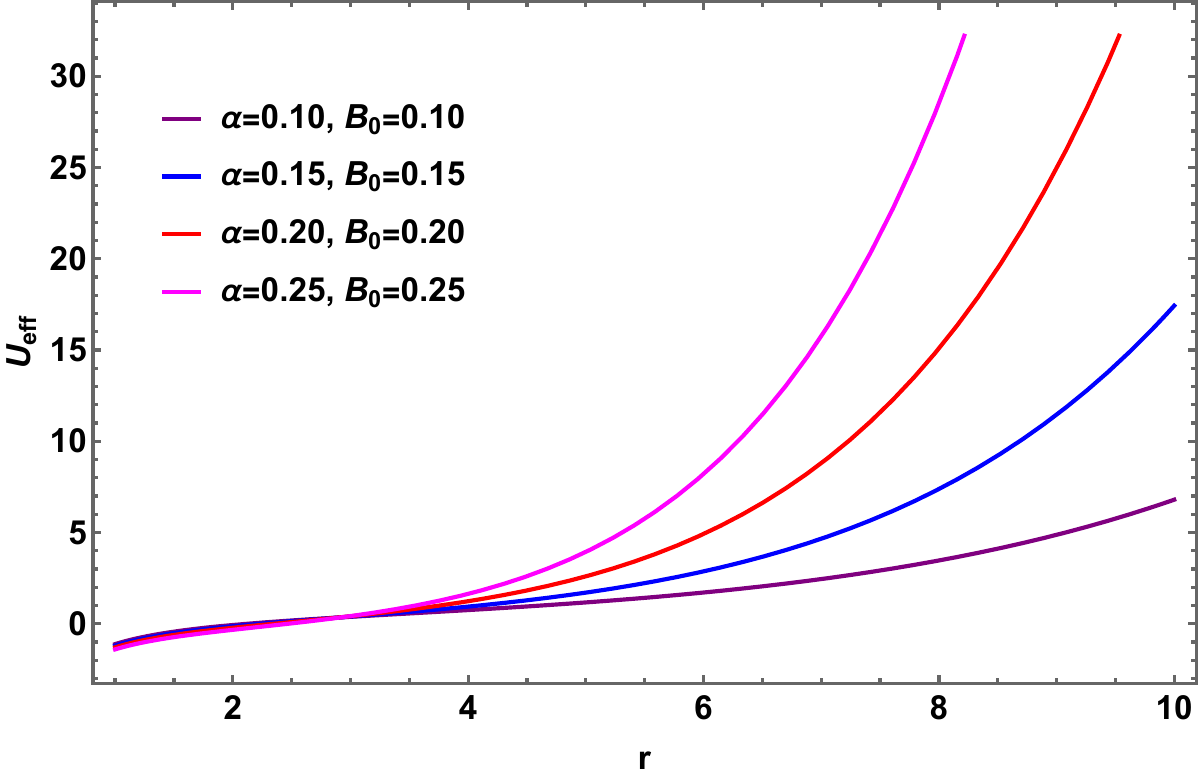}} 
    \caption{The behavior of the effective potential $U_\text{eff}(r)$ for neutral particles by varying $\alpha$ and $B_0$. Here $M=1$, $\mathcal{L}_0=0.01$, and $\ell_p=10$.}
    \label{fig:potential33}
\end{figure}

\begin{figure}[ht!]
    \centering
    \includegraphics[width=0.5\linewidth]{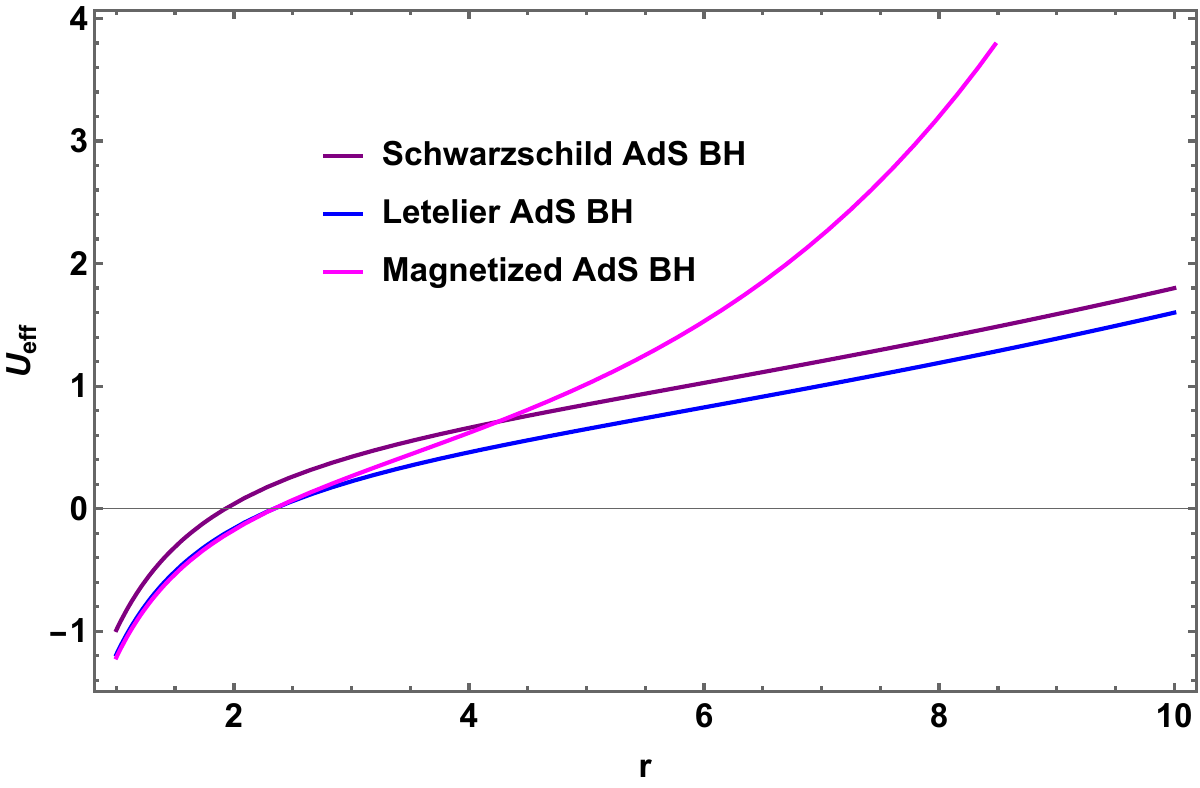}
    \caption{A comparison of the effective potential $U_\text{eff}(r)$ for neutral particles under BH scenario. Here $M=1$, $\mathcal{L}_0=0.01$, and $\ell_p=10$. Purple color: $\alpha=0, B_0=0$; blue color: $\alpha=0.1$, $B_0=0.1$; magenta: $\alpha=0.2, B_0=0.1$.}
    \label{fig:potential44}
\end{figure}

Figure~\ref{fig:potential33} illustrates how the effective potential \( U_\text{eff}(r) \) varies with radial distance \( r \) for different values of the CS parameter \( \alpha \) and magnetic field strength \( B_0 \). In panel (a), increasing \( \alpha \) causes the effective potential to gradually shift downward, suggesting that higher values of the CS parameter influence and potentially control the dynamics of neutral particles. In contrast, panel (b) shows that as \( B_0 \) increases, the effective potential shifts upward, indicating that stronger magnetic fields enhance the particle motion within the gravitational field. Panel (c) demonstrates a similar upward trend when both \( \alpha \) and \( B_0 \) are increased simultaneously, implying that the combined effects of the CS parameter and magnetic field significantly alter the motion of particles.

In Figure \ref{fig:potential44}, we present a comparison of the effective potential $U_\text{eff}(r)$ under different BH scenario: the Schwarzschild AdS BH ($\alpha=0=B_0$); the Letelier AdS BH ($\alpha=0.2,B_0=0$), and magnetized AdS BH ($\alpha=0.2,B_0=0.1$) keeping other parameters fixed $M = 1; \ell_p= 10; \mathcal{L}_0= 0.01$.

One can study motion of neutral particle in circular orbits for which the conditions $\dot{r}=0$ and $\ddot{r}=0$ must satisfied. These conditions implies the following two relations:
\begin{eqnarray}
    &&\mathcal{E}^2=U_\text{eff}(r),\label{ss8}\\
    &&U'_\text{eff}(r)=0.\label{ss9}
\end{eqnarray}

Simplification of these relations results the following physical quantities of time-like neutral particle given by 
\begin{eqnarray}
    &&\mathcal{L}^2_0=\frac{r^3}{\tilde{\Lambda}^2}\,\frac{2\,(\tilde{\Lambda})'\,\mathcal{F}+\tilde{\Lambda}\,\mathcal{F}'}{(2\,\mathcal{F}-r\,\mathcal{F}')\,\tilde{\Lambda}-4\,r\,(\tilde{\Lambda})'\,\mathcal{F}},\label{ss10}\\
    &&\mathcal{E}^2=\frac{2\,\tilde{\Lambda}^2\,\mathcal{F}^2\,\left(\tilde{\Lambda}-r\,(\tilde{\Lambda})'\right)}{(2\,\mathcal{F}-r\,\mathcal{F}')\,\tilde{\Lambda}-4\,r\,(\tilde{\Lambda})'\,\mathcal{F}},\label{ss11}
\end{eqnarray}
where prime denotes ordinary derivative w. r. t. $r$.

Substituting the metric function $\mathcal{F}$ and $\tilde{\Lambda}$ and after simplification results:
\begin{eqnarray}
    &&\mathcal{L}_0=\sqrt{\frac{r^3}{(1+B^2_0\,r^2)^2}}\,\sqrt{\frac{r^3 + 3\,B^2_0\, r^5 + \ell_p^2\,\left(M - 3\, B^2_0\, M\, r^2 - 2\, B^2_0\, r^3\, (-1 + \alpha)\right)}{-4\,B^2_0\, r^6 + \ell_p^2\, r\, \left[M\, (-3 + 5\, B^2_0\, r^2) + r\, (-1 + 3\, B^2_0\, r^2)\,(-1 + \alpha)\right]}},\label{ss12}\\
    &&\mathcal{E}=\pm\,\sqrt{\frac{(1 - B^2_0\, r^2)\, (1 + B^2_0\,r^2)^2\,\left(r^3+ \ell_p^2\, (-2\, M + r - r\,\alpha)\right)^2}{-4\,B^2_0\, r^6 + \ell_p^2\, r\, \left[M\, (-3 + 5\, B^2_0\, r^2) + r\, (-1 + 3\, B^2_0\, r^2)\,(-1 + \alpha)\right]}}.\label{ss13}
\end{eqnarray}

From expressions~(\ref{ss12}) and~(\ref{ss13}), it is evident that the specific angular momentum and energy per unit mass of time-like neutral particles in circular orbits around the BH are affected by several key parameters. These include the cosmic string parameter $\alpha$, the curvature radius $\ell_p$, the BH mass $M$, and the magnetic field strength $B_0$.

\begin{figure}[ht!]
    \centering
    \subfloat[$B_0=0.01$]{\centering{}\includegraphics[width=0.45\linewidth]{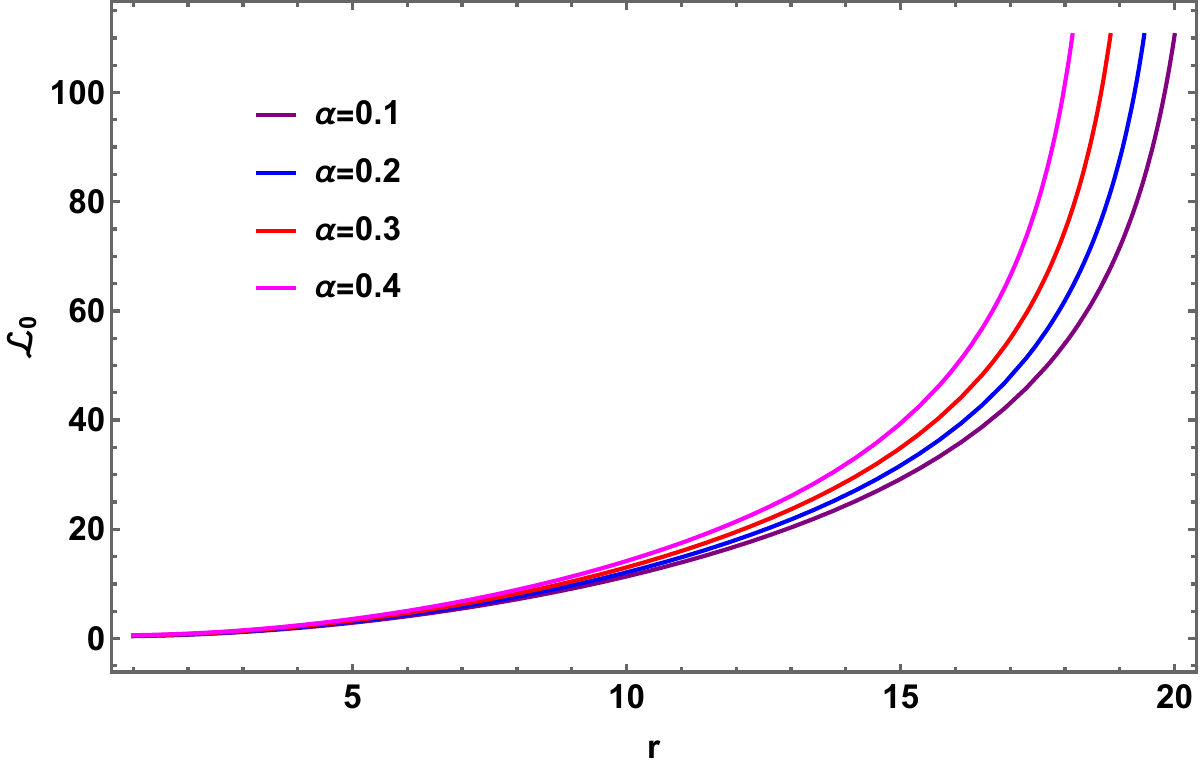}}\quad\quad
    \subfloat[$B_0=0.01$]{\centering{}\includegraphics[width=0.45\linewidth]{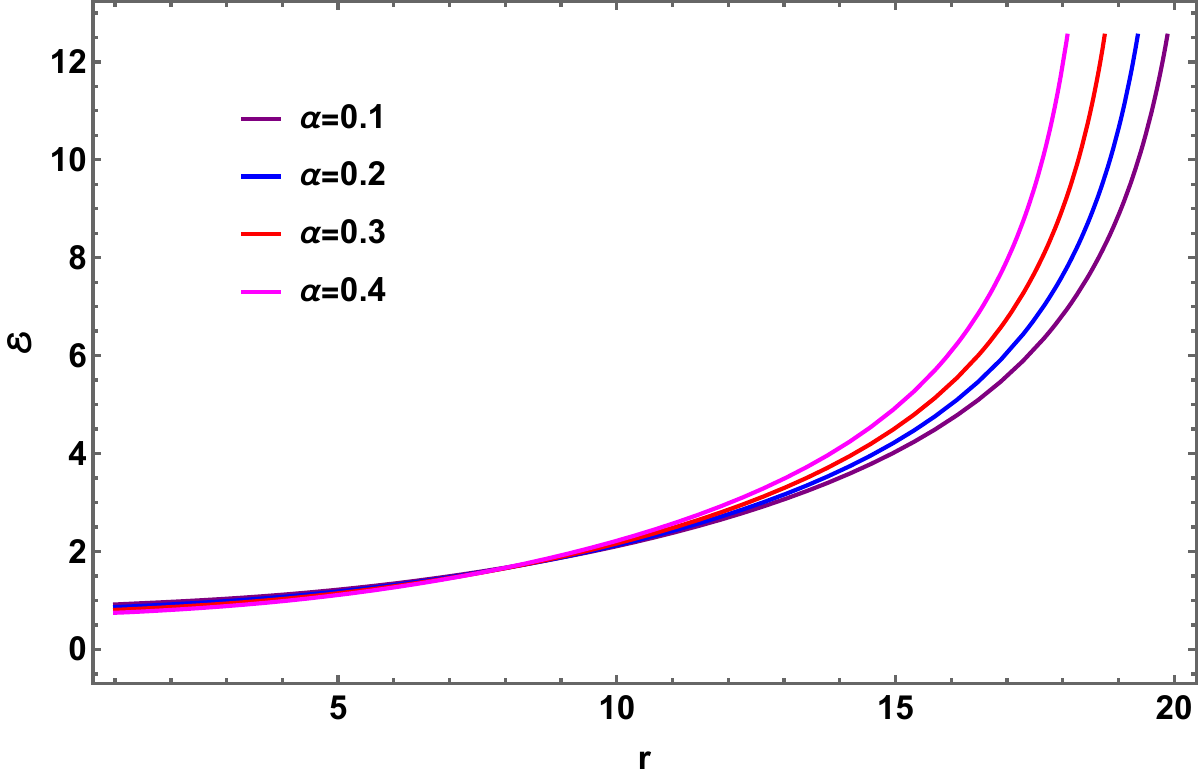}}\\
    \subfloat[$\alpha=0.1$]{\centering{}\includegraphics[width=0.45\linewidth]{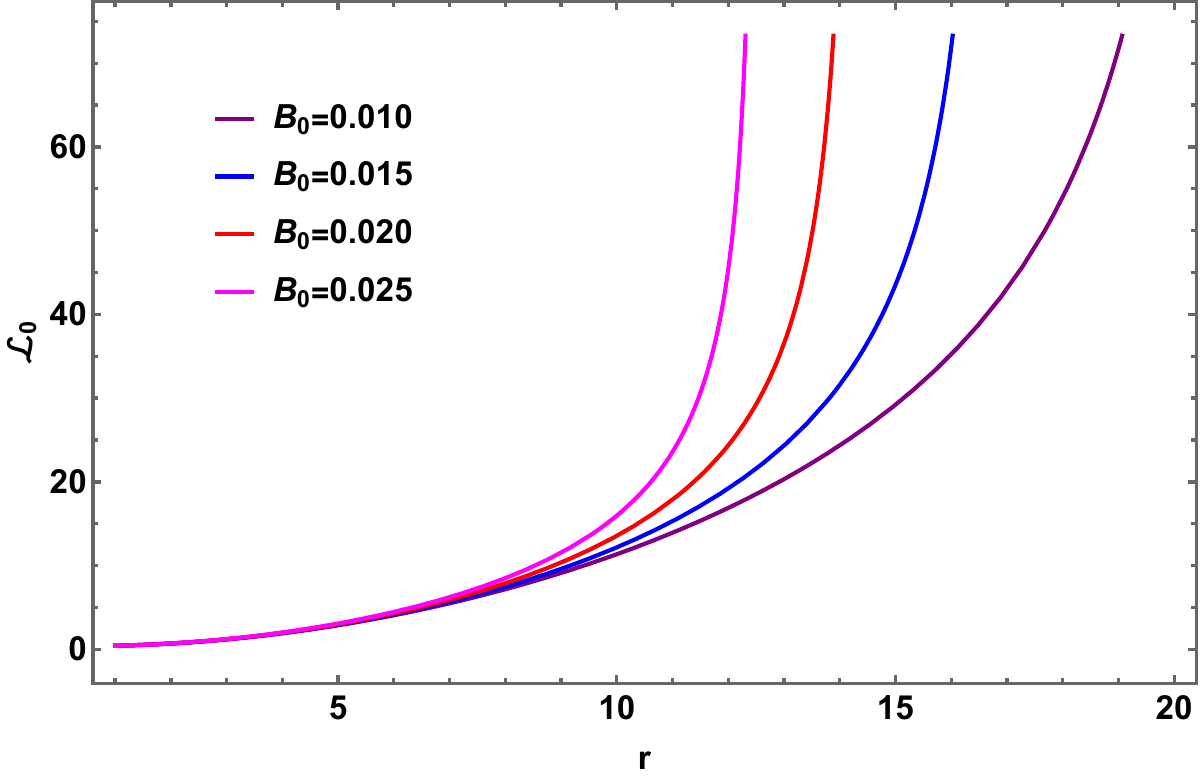}}\quad\quad
    \subfloat[$\alpha=0.1$]{\centering{}\includegraphics[width=0.45\linewidth]{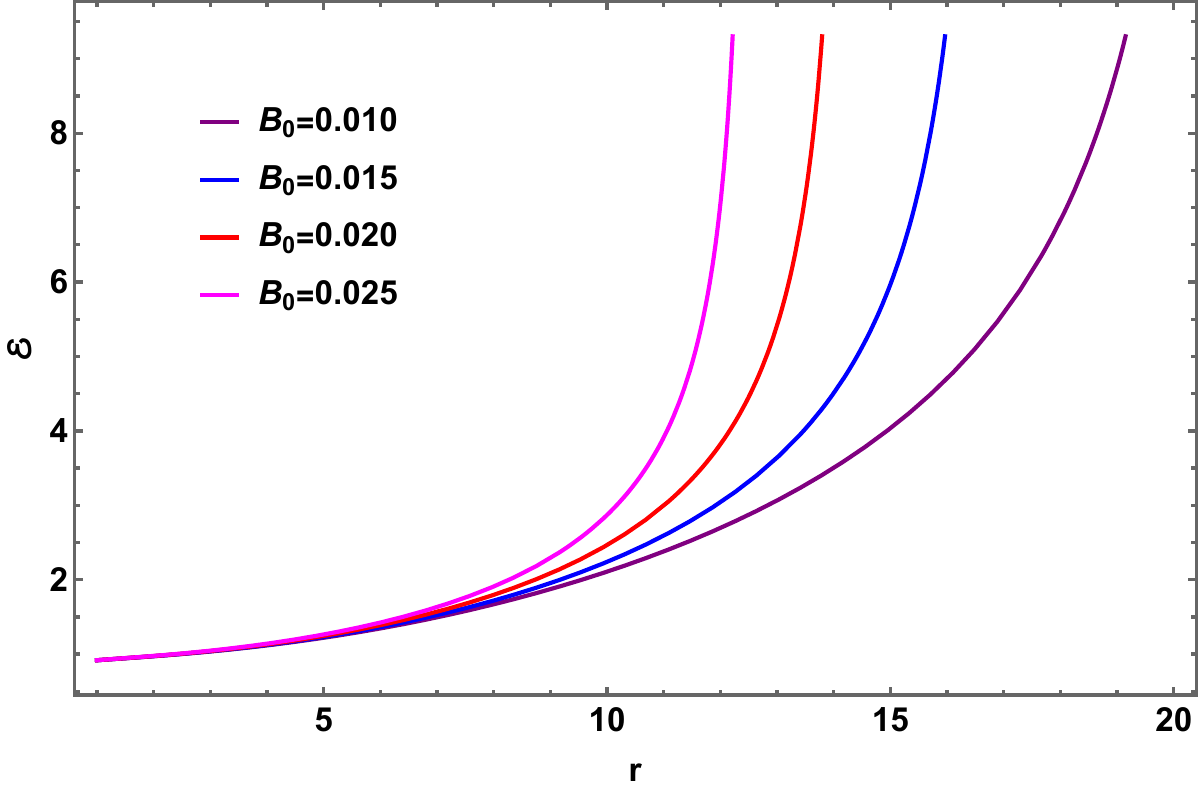}}\\
    \subfloat[]{\centering{}\includegraphics[width=0.45\linewidth]{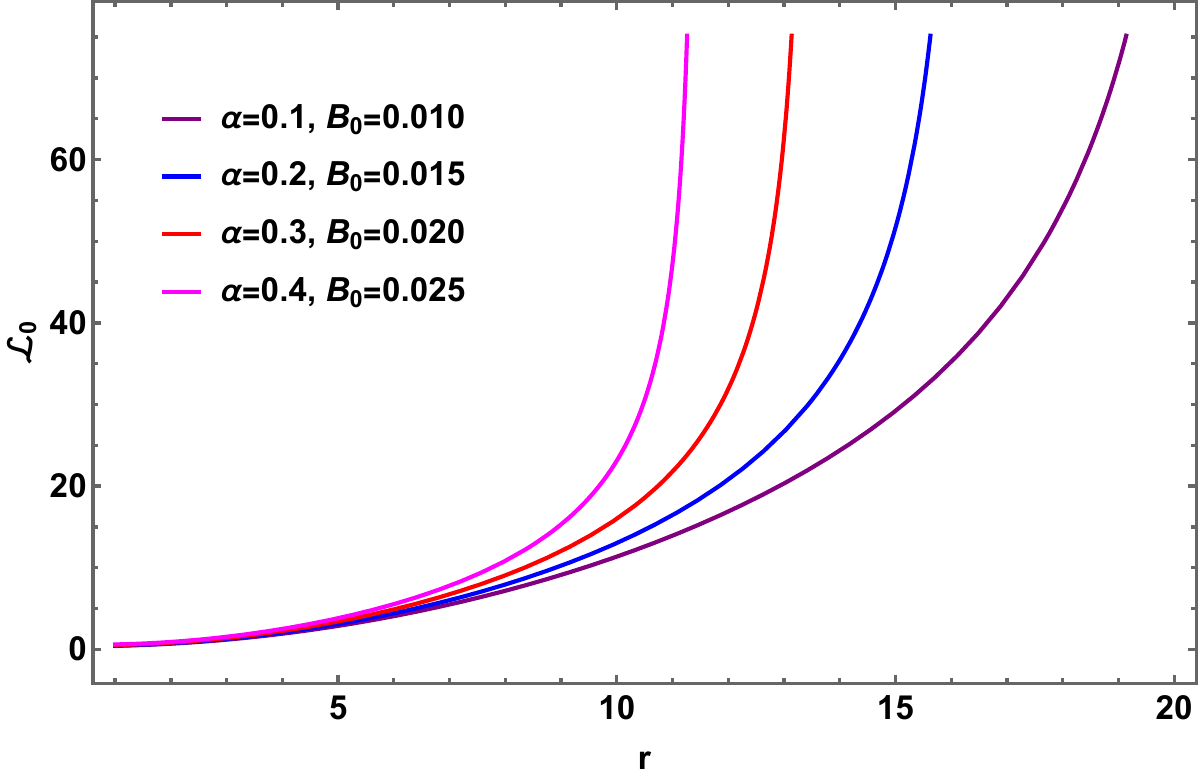}}\quad\quad
    \subfloat[]{\centering{}\includegraphics[width=0.45\linewidth]{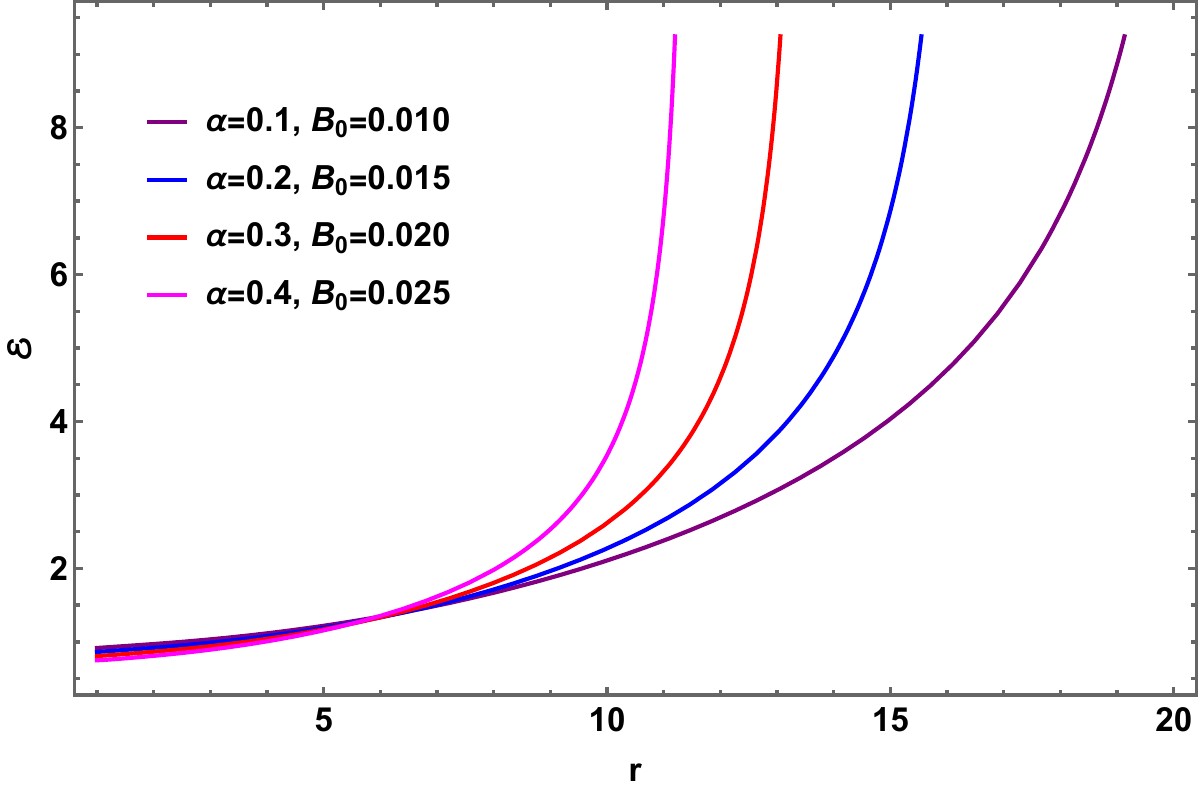}}
    \caption{The behavior of the specific angular momentum $\mathcal{L}_0$ (left column) and energy $\mathcal{E}$ (right column) of time-like neutral particle orbiting in circular paths for different values of the CS parameter $\alpha$ and the magnetic field strength $B_0$. Here, we set $M=0.1$ and the curvature radius $\ell_p=10$.}
    \label{fig:neutral-particle}
\end{figure}

In Figure~\ref{fig:neutral-particle}, we present a series of plots illustrating the variation of the specific angular momentum per unit mass $\mathcal{L}_0$ and the specific energy per unit mass $\mathcal{E}$ as functions of the radial distance $r$, for different values of the cosmic string parameter $\alpha$ and the magnetic field strength $B_0$. The left column corresponds to $\mathcal{L}_0$, while the right column shows $\mathcal{E}$. From all the panels in the left column, it is evident that the specific angular momentum increases with an increase in either the cosmic string parameter $\alpha$, the magnetic field strength $B_0$, or both. A similar trend is observed for the specific energy in the right column.

In the limit $B_0=0$, corresponding to the absence of the magnetic field, the selected BH space-time reduces to the Letelier AdS space-time. In this limit, the specific angular momentum and energy of time-like neutral particles from Eqs. (\ref{ss12})-(\ref{ss13}) reduce as
\begin{eqnarray}
    &&\mathcal{L}_0=r\,\sqrt{\frac{\frac{M}{r}+\frac{r^2}{\ell^2_p}}{1-\alpha-\frac{3\,M}{r}}},\label{ss14}\\
    &&\mathcal{E}=\pm\,\frac{1-\alpha-\frac{2\,M}{r}+\frac{r^2}{\ell^2_p}}{\sqrt{1- \alpha-\frac{3\,M}{r}}}.\label{ss15}
\end{eqnarray}

\subsection{Determining the innermost stable circular orbits}

The minimum and maximum values of the effective potential correspond to stable and unstable circular orbits, respectively. In Newtonian gravity, the innermost stable circular orbit (ISCO) does not have a minimum bound on the radius; the effective potential always exhibits a minimum for any given angular momentum. However, this behavior changes when the effective potential depends not only on the angular momentum of the particle but also on additional factors, such as spacetime curvature or external fields.

In GR, the effective potential for particles orbiting near a Schwarzschild BH features two extrema-one minimum and one maximum-for a given angular momentum. At \( r = 3\,r_s \), where \( r_s \) is the Schwarzschild radius, these two extrema coincide at a critical value of angular momentum, marking the location of the ISCO. This point defines the transition from stable to unstable circular orbits. The ISCO can be determined by applying the following conditions to the effective potential \( U_\text{eff}(r) \):
\begin{align}
U_\text{eff}&=\mathcal{E}^2, \\
\frac{dU_\text{eff}}{dr} &= 0 \quad \text{(circular orbit condition)}, \\
\frac{d^2U_\text{eff}}{dr^2} &= 0 \quad \text{(marginal stability condition)}.
\end{align}
Using the effective potential expression given in Eq. (\ref{ss7}), one can determine the position of ISCO. The expression of the last equation is too long which we omitted it. 

\subsection{Effective force dynamics in BH spacetime}

The effective force acting on a particle governs its motion in the gravitational field of a BH, indicating whether the particle is attracted toward or repelled from the BH. In this study, we investigate the motion of particles in a magnetized AdS BH  background where both attractive and repulsive forces may arise, depending on the values of the system parameters-such as the cosmic string and magnetic field strength. This highlights the critical role of the effective force in determining the stability and dynamics of particle trajectories.

In this context, we compute the effective force acting on the particles using equation~(\ref{ss6}), which is given by:
\cite{AB1,AB2,AB3,AB4,AB5,AB6}
\begin{eqnarray}
    &&\mathcal{F}_0=-\frac{1}{2}\,\frac{\partial U_\text{eff}}{\partial r}=-\frac{(1+B^2_0\,r^2)}{r^4\,\ell^2_p}\,\Big[(1 + B^2_0\, r^2)\left\{r^2 +\mathcal{L}^2_0\,(1+B^2_0\,\, r^2)^2\right\}(r^3 +M\, \ell^2_p)+\mathcal{L}_0^2(1 + B^2_0\, r^2)(-1 + B^4_0\,r^4)\times\nonumber\\
    &&\left\{r^3 - 2\, M\,\ell^2_p-r\,\ell^2_p (-1 + \alpha)\right\}+2\,B^2_0\,r^2\,\left\{r^2 + (\mathcal{L}_0 + B^2_0\,\mathcal{L}_0\, r^2)^2\right\}\left\{r^3 - 2\, M\, \ell^2_p -r\,\ell^2_p(-1 + \alpha)\right\}\Big].\label{ss16}
\end{eqnarray}

From the above expression (\ref{ss16}), it is clear that the force on neutral particles in the gravitational field is influenced by several factors. These include the CS parameter $\alpha$, the magnetic field strength $B_0$, the BH mass $M$, the specific angular momentum per unit mass $\mathcal{L}_0$, and the curvature radius $\ell_p$.

\begin{figure}[ht!]
    \centering
    \subfloat[$B_0=0.1$]{\centering{}\includegraphics[width=0.45\linewidth]{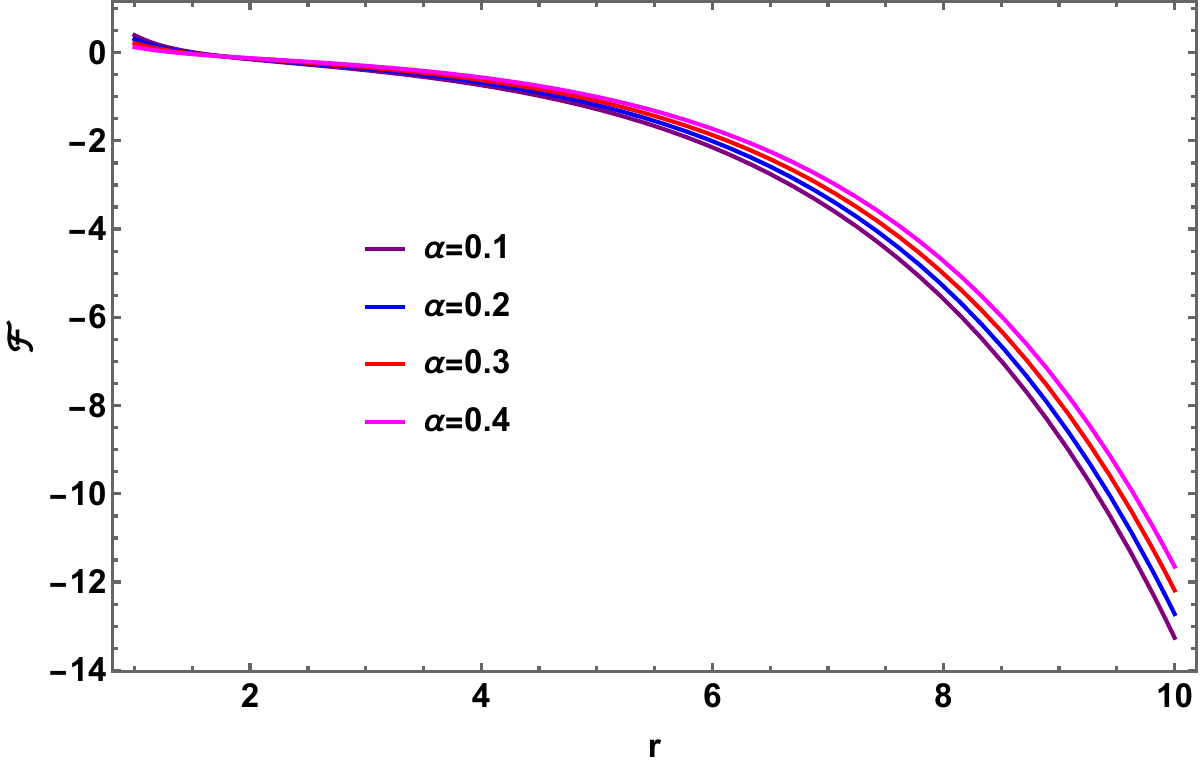}}\quad\quad
    \subfloat[$\alpha=0.1$]{\centering{}\includegraphics[width=0.45\linewidth]{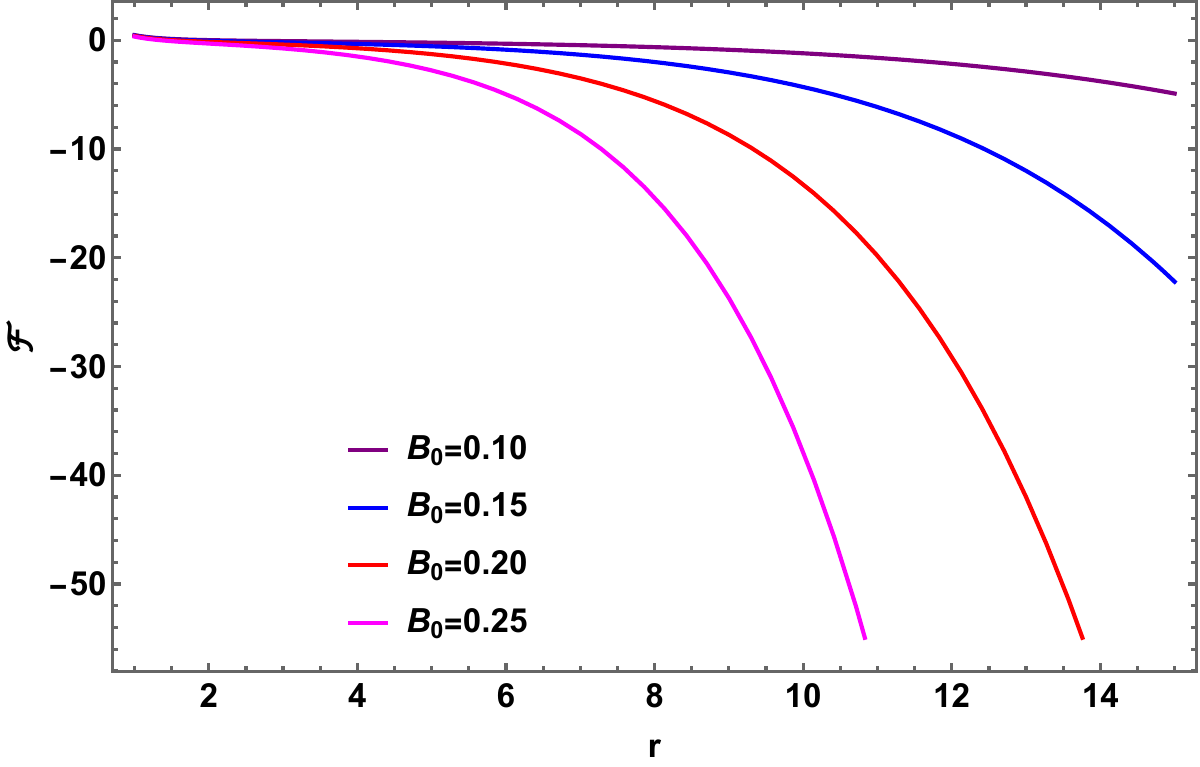}}\\
    \subfloat[]{\centering{}\includegraphics[width=0.45\linewidth]{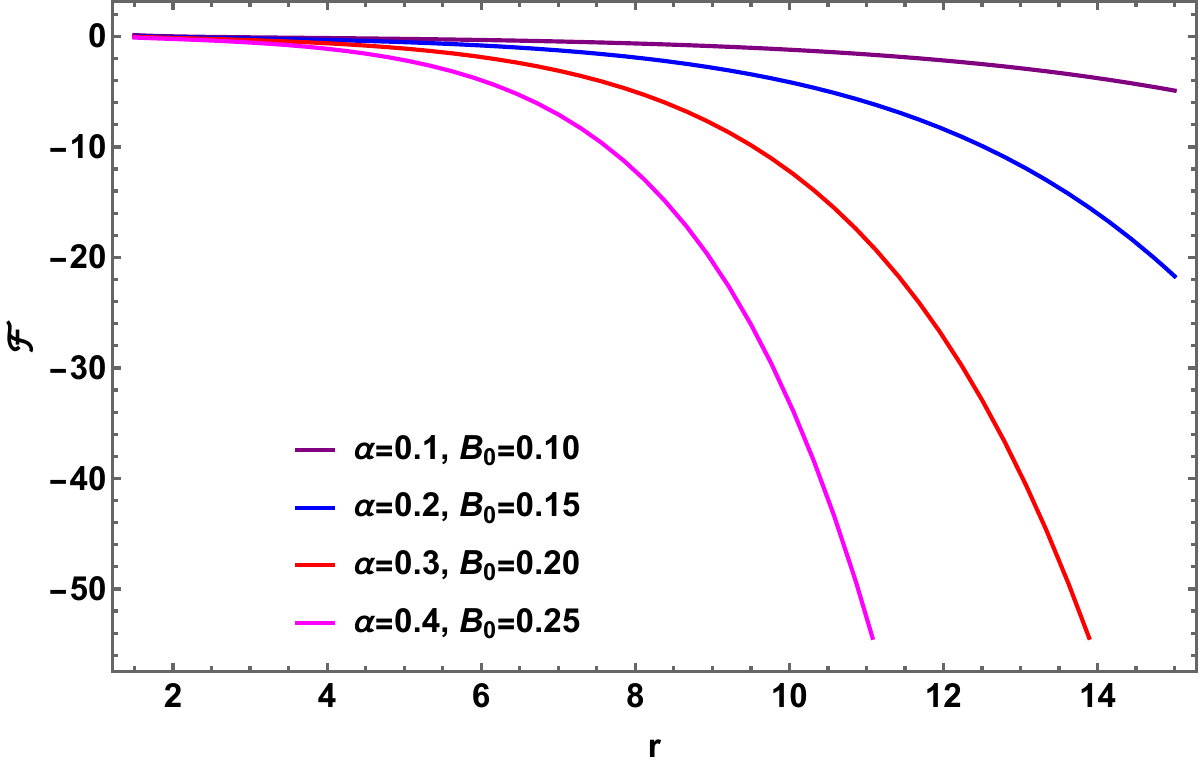}} 
    \caption{The behavior of force $\mathcal{F}_0(r)$ by varying $\alpha$ and $B_0$. Here $M=0.1$, $\mathcal{L}_0=1$, and $\ell_p=10$.}
    \label{fig:force11}
\end{figure}

\begin{figure}[ht!]
    \centering
    \includegraphics[width=0.5\linewidth]{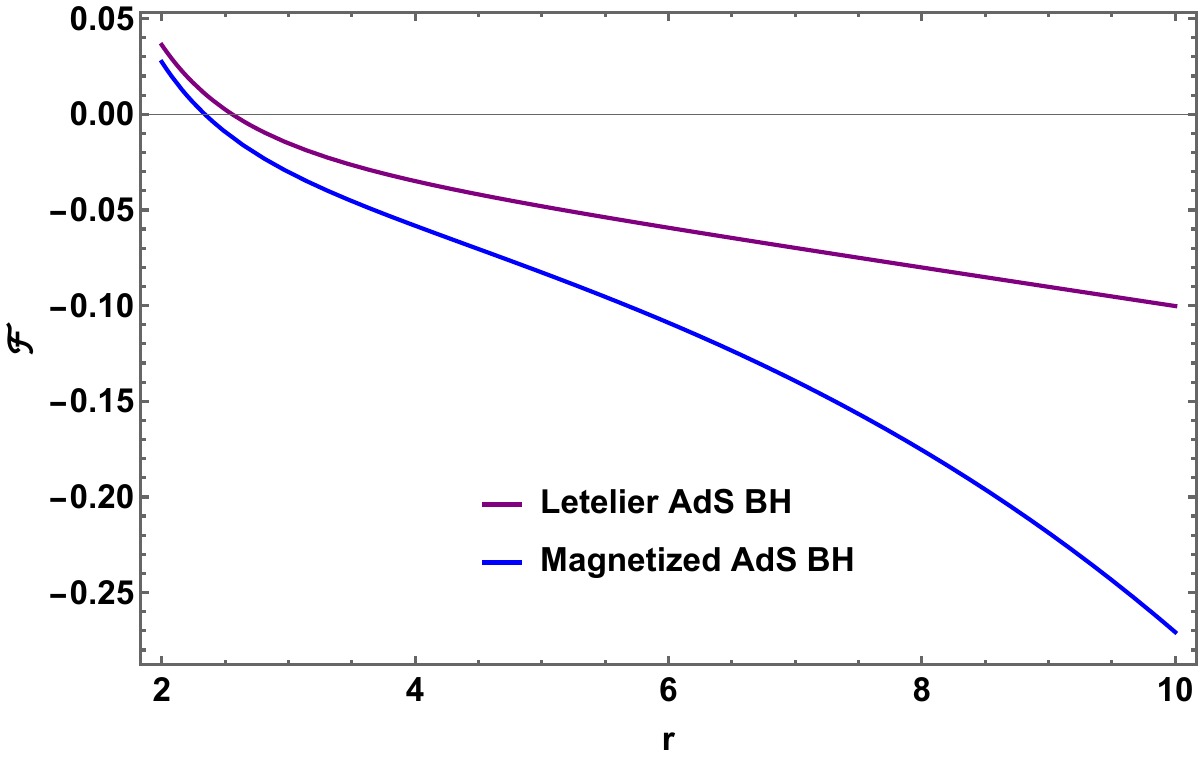}
    \caption{A comparison of of force $\mathcal{F}_0(r)$ under different BH scenario. Here $M=0.1$, $\mathcal{L}_0=1$, and $\ell_p=10$. Purple color: $\alpha=0.2$, $B_0=0$; blue: $\alpha=0.2, B_0=0.05$.}
    \label{fig:force22}
\end{figure}

Figure~\ref{fig:force11} illustrates the variation of the force \( \mathcal{F}_0(r) \) with radial distance \( r \) for different values of the CS parameter \( \alpha \) and magnetic field strength \( B_0 \). Across all panels, it is observed that the magnitude of the negative force on particles increases gradually as \( r \) increases, for fixed values of \( \alpha \), \( B_0 \), or their combination. This behavior suggests that both the CS parameter and the magnetic field-either independently or jointly-reduce the effective force acting on neutral particles, making the interaction more attractive in nature. In other words, the attractive force on neutral particles becomes stronger at larger radial distances.

In Figure~\ref{fig:force22}, we present a comparative analysis of the force \( \mathcal{F}_0(r) \) under two different BH configurations: the Letelier AdS BH (with \( \alpha = 0.2 \), \( B_0 = 0 \)) and the magnetized AdS BH (with \( \alpha = 0.2 \), \( B_0 = 0.05 \)). Other parameters are kept fixed at \( M = 0.1 \), \( \ell_p = 10 \), and \( \mathcal{L}_0 = 1 \). The comparison shows that the inclusion of a magnetic field enhances the attractive nature of the force, indicating that the magnetic field has a significant influence on the motion of neutral particles in the BH background.

\section{Optical properties in magnetized Letelier AdS spacetime} \label{isec5}

In this section, we examine the optical properties of the magnetized Letelier BH using the metric (\ref{bb1}) and explore how different parameters of the space-time geometry affect the motion of photon particles near the BH. The study of null geodesic motion provides crucial information about the behavior of photon particles, including the photon sphere radius, the BH's shadow, photon trajectories, the stability of circular orbit, and the period of this orbit. A few recent studies of optical properties or null geodesics analysis in various BH were reported in Refs. \cite{AB3,AB4,AB5,NPB,CJPHY,AHEP1,AHEP2,EPJC,AHEP3}. 

We consider the null geodesic motion in the equatorial plane defined by $\theta=\pi/2$ and $\dot{\theta}=0$. Using the condition $ds=0$ for the metric (\ref{bb1}) yields
\begin{equation}
    \tilde{\Lambda}^2\, \left(-\mathcal{F}(r)\,\dot{t}^2+\frac{\dot{r}^2}{\mathcal{F}(r)} \right)+\frac{r^2}{\tilde{\Lambda}^2}\,\dot{\phi}^2=0,\label{dd1}
\end{equation}

As stated earlier, the space-time is a static and spherically symmetric, and hence, there are two conserved quantities known as the energy ($\mathrm{E}$) and angular momentum ($\mathrm{L}$). These are given by
\begin{equation}
    \dot{t}=\frac{\mathrm{E}}{\tilde{\Lambda}^2\,\mathcal{F}},\quad\quad\quad \dot{\phi}=\frac{\tilde{\Lambda}^2\,\mathrm{L}}{r^2},\quad\quad\quad \bar{\Lambda}=1+B^2_0\,r^2.\label{dd2}
\end{equation}

With these, the geodesics equation for $r$ coordinate from Eq. (\ref{dd1}) becomes
\begin{equation}
    \dot{r}^2=\frac{\mathrm{E}^2}{\tilde{\Lambda}^2}-\frac{\mathrm{L}^2}{r^2}\,\mathcal{F}.\label{dd3}
\end{equation}
The above equation can be re-written as
\begin{equation}
    \tilde{\Lambda}^4\,\dot{r}^2+V^\text{null}_\text{eff}(r)=\mathrm{E}^2,\label{dd3aa}
\end{equation}
where the effective potential is given by
\begin{equation}
    V^\text{null}_\text{eff}(r)=\frac{\mathrm{L}^2}{r^2}\,\tilde{\Lambda}^4\,\mathcal{F}=\frac{\mathrm{L}^2}{r^2}\,(1+B^2_0\,r^2)^4\,\left(1-\alpha-\frac{2\,M}{r}+\frac{r^2}{\ell^2_p}\right).\label{dd3bb}
\end{equation}

From the expression in Eq. (\ref{dd3bb}), it is clear that the effective potential for null geodesics motion is influenced by several parameters that are present in the BH space-time. These include the magnetic field strength $B_0$, the CS parameter $\alpha$, the radius of curvature $\ell_p$. Moreover, the effective potential gets modification by the BH mass $M$ and the angular momentum $\mathrm{L}$.

\begin{figure}[ht!]
    \centering
    \subfloat[$B_0=0.1,\,\mathrm{L}=1$]{\centering{}\includegraphics[width=0.45\linewidth]{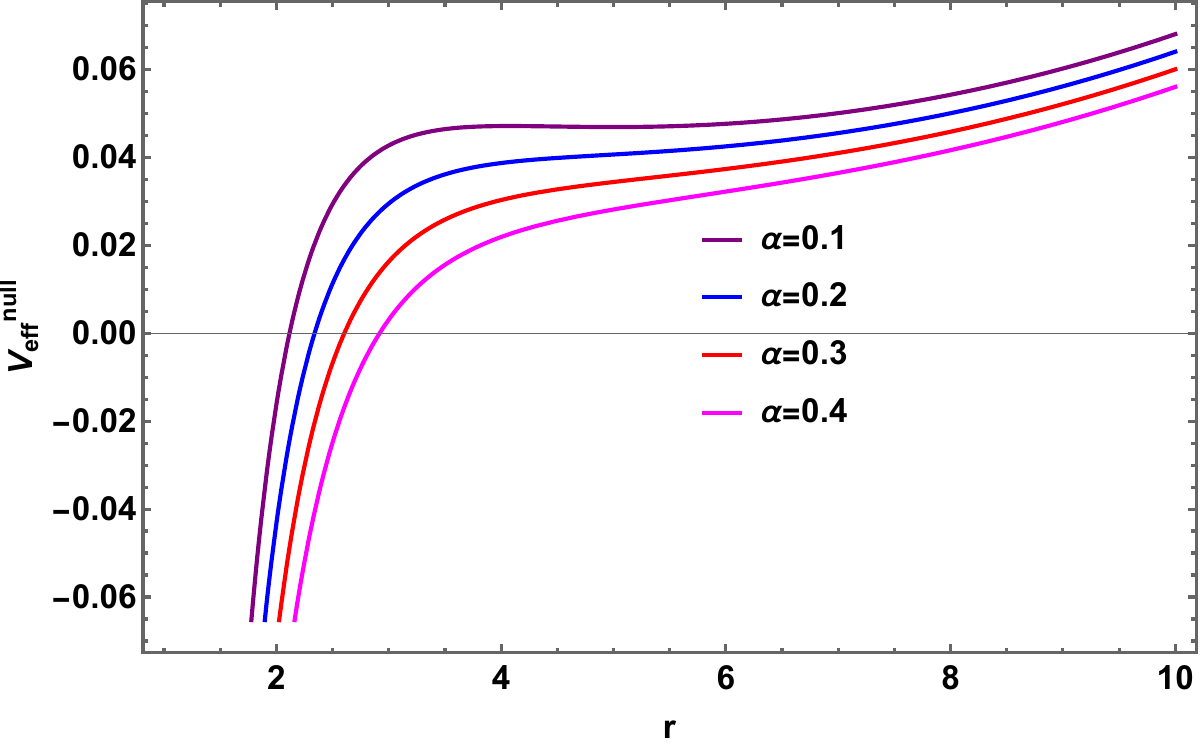}}\quad\quad
    \subfloat[$\alpha=0.1,\, \mathrm{L}=1$]{\centering{}\includegraphics[width=0.45\linewidth]{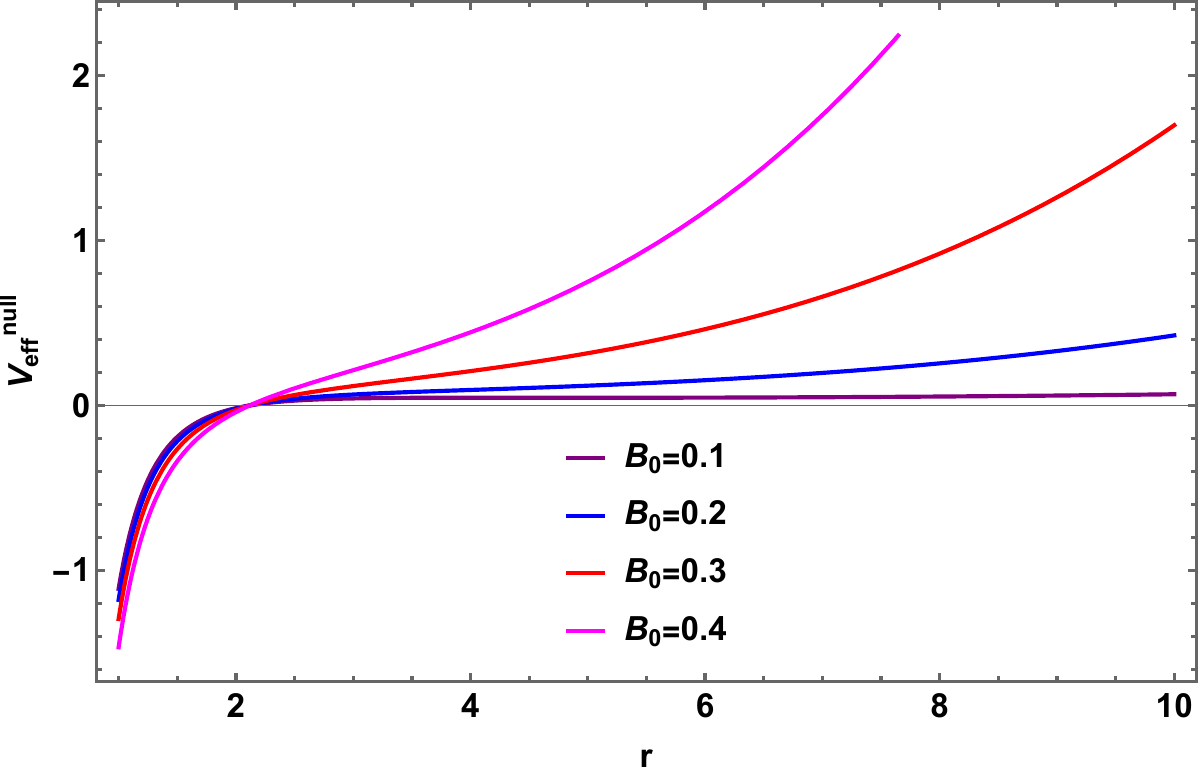}}\\
    \subfloat[$\mathrm{L}=1$]{\centering{}\includegraphics[width=0.45\linewidth]{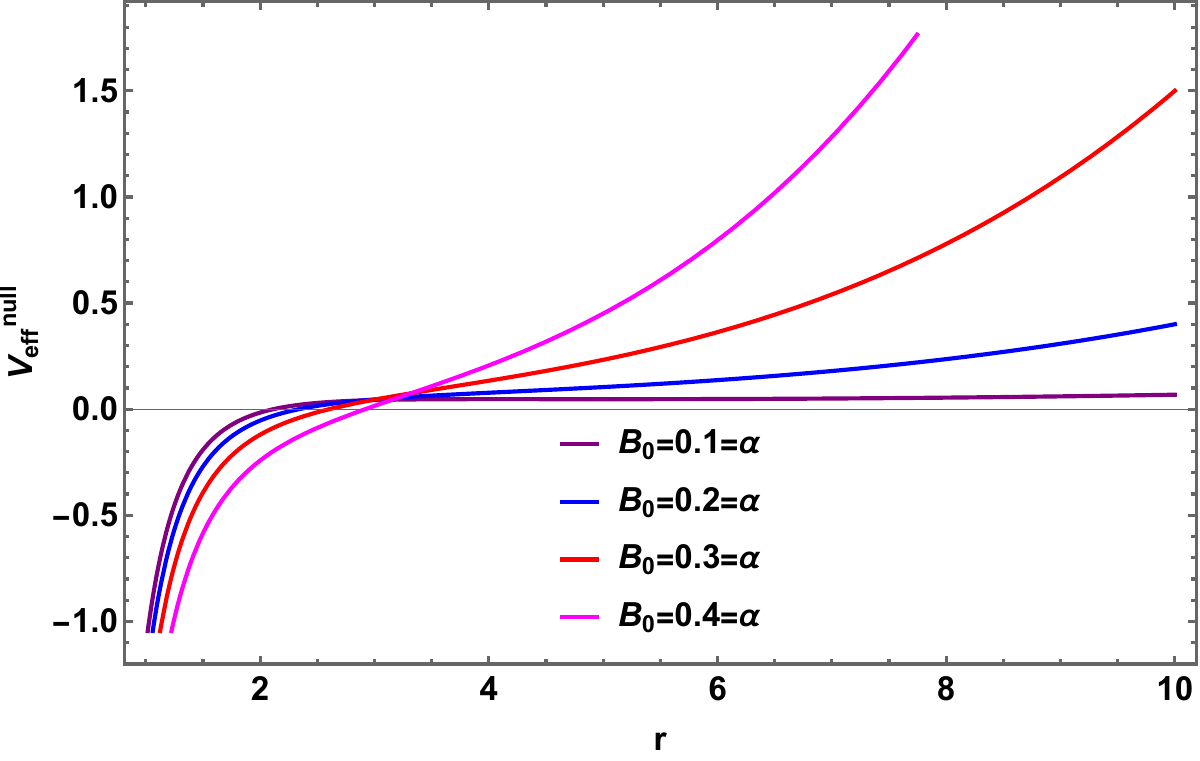}}\quad\quad
    \subfloat[$B_0=0.1=\alpha$]{\centering{}\includegraphics[width=0.45\linewidth]{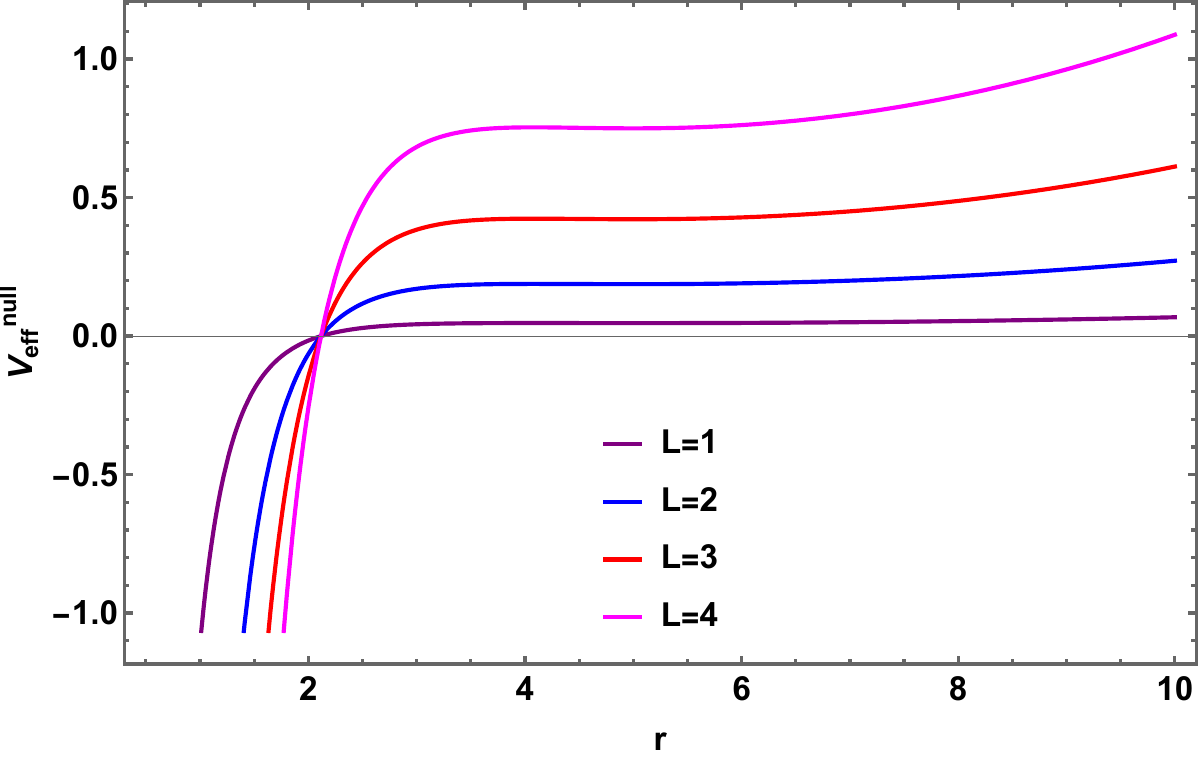}}
    \caption{The behavior of the effective potential for null geodesics by varying $\alpha$, $B_0$ and $\mathrm{L}$. Here $M=1$, abd $\ell_p=10$.}
    \label{fig:potential3}
\end{figure}

\begin{figure}[ht!]
    \centering
    \includegraphics[width=0.5\linewidth]{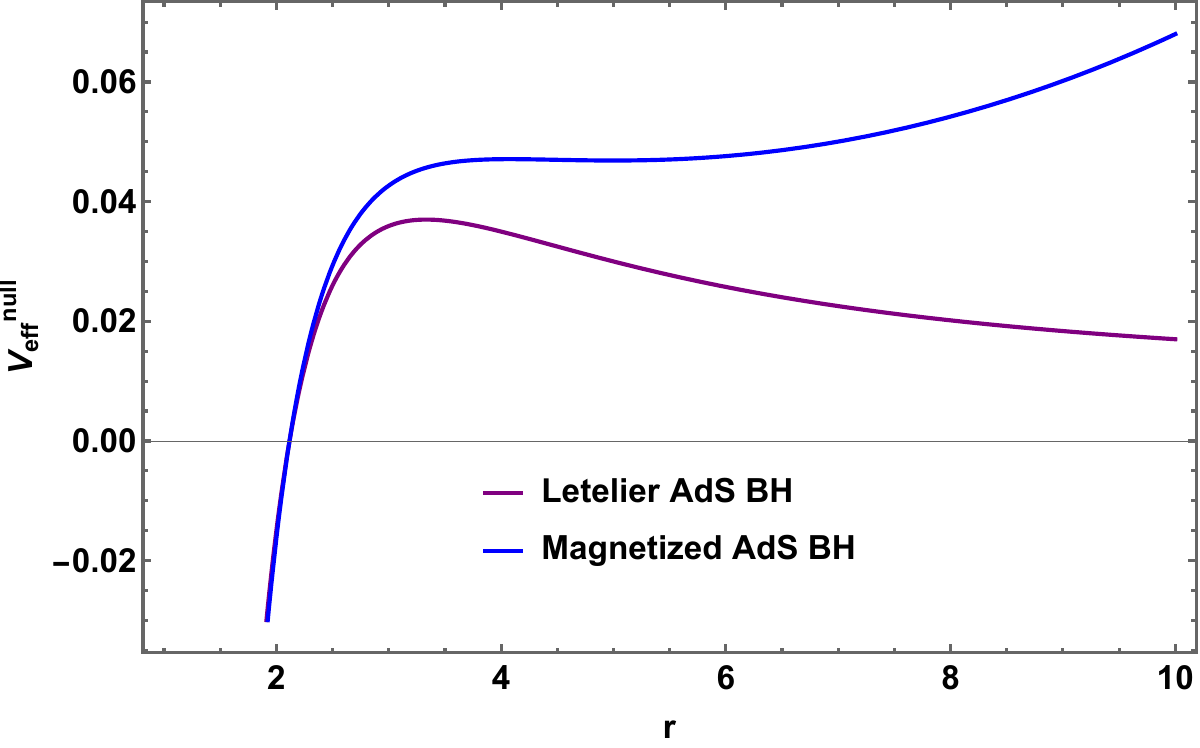}
    \caption{A comparison of the effective potential for null geodesics. Here $M=1$, $\mathrm{L}=1$, and $\ell_p=10$. Purple color: $\alpha=0.1, B_0=0$, blue color: $B_0=0.1, \alpha=0.1$.}
    \label{fig:potential4}
\end{figure}

In Fig. \ref{fig:potential3}, we present the behavior of the effective potential for null geodesics under variations of the cosmic string (CS) parameter $\alpha$, the magnetic field strength $B_0$, and the angular momentum $\mathrm{L}$. Panel (a) shows that increasing the CS parameter from $\alpha = 0.1$ leads to a decrease in the effective potential as a function of the radial coordinate $r$. This behavior suggests that larger values of $\alpha$ keeping the magnetic field strength $B_0$ fix effectively weaken the influence of the gravitational field, thereby reducing its effect on photon particles.

In contrast, panels (b) to (d) demonstrate that slight increases in either $B_0$ and $\mathrm{L}$, or in the combined values of ($\alpha$, $B_0$), result in an increase in the effective potential with increasing radial distance. This indicates that the combined influence of the CS parameter $\alpha$ and the magnetic field $B_0$ strengthens the gravitational field generated by the magnetized BH, thus enhancing the effective potential experienced by the photon particles. In this figure, the BH mass is fixed at $M = 1$, and the AdS radius at $\ell_p = 10$. These parameters collectively govern the dynamics of photon particles and influence whether they are captured by or escape from the gravitational field of the BH.

In Fig. \ref{fig:potential4}, we present a comparison of the effective potential for null geodesics with and without the influence of a magnetic field. It is observed that a slight increase in the magnetic field strength, from $B_0 = 0$ to $B_0 = 0.1$, leads to a more pronounced variation in the effective potential of the null geodesics. In this figure, the BH mass is fixed at $M = 1$, the angular momentum $\mathrm{L}=1$, and the AdS radius at $\ell_p = 10$.

Now, we focus on circular orbits motion and discuss the relevant quantities associated with these. For circular orbits of radius $r=r_c$, we have the conditions $\dot{r}=0$ and $\ddot{r}=0$. Thereby, using (\ref{dd3aa}), we find the following two relations
\begin{equation}
    V_\text{eff}(r)=\mathrm{E}^2,\quad\quad V'_\text{eff}(r)=0.\label{dd3cc}
\end{equation}

The first relation $V_\text{eff}(r)=\mathrm{E}^2$ gives us the critical impact parameter for photon particle and is given by
\begin{equation}
    \frac{1}{\beta_c}=\frac{\mathrm{E}}{\mathrm{L}}=\frac{(1+B^2_0\,r^2)^2}{r}\sqrt{1-\alpha-\frac{2\,M}{r}+\frac{r^2}{\ell^2_p}}.\label{dd3dd}
\end{equation}

From expression given in Eq. (\ref{dd3dd}), it is evident that the impact parameter for photon particles originating from infinity and reaching a minimum distance to turn back from the BH is influenced by several factors. These include the magnetic field strength $B_0$, the CS parameter $\alpha$, and the AdS radius $\ell_p$. Additionally, the BH mass $M$ alters the impact parameter. 

The second relation $V'_\text{eff}(r)=0$ gives us the photon sphere radius $r=r_\text{ph}$ given by the following equation
\begin{equation}
    3\,M-(1-\alpha)\,r-5\,M\,B^2_0\,r^2+3\,B^2_0\,(1-\alpha)\,r^3+\frac{4\,B^2_0}{\ell^2_p}\,r^5=0.\label{dd3ee}
\end{equation}
The exact real-valued expression for the photon sphere radius is challenging to determine analytically. However, it can be computed numerically by assigning appropriate values to the magnetic field strength $B_0$, the CS parameter $\alpha$, the BH mass $M$ and the AdS radius $\ell_p$. 

In the scenario where $B_0=0$, meaning there is no magnetic field, the photon sphere of the Letelier BH AdS metric reduces to
\begin{equation}
    r=r_\text{ph}=\frac{3\,M}{1-\alpha}>r_\text{sch}(=3\,M).\label{dd3ff}
\end{equation}

We now aim to determine photon trajectories in the gravitational field and examine how various parameters influence the motion of photon particles as they move under the influence of the gravitational field generated by the magnetized BH. From Eqs. (\ref{dd2}) and (\ref{dd3}), we defined the following equation
\begin{equation}
    \left(\frac{dr}{d\phi}\right)^2=\frac{\dot{r}^2}{\dot{\phi}^2}=\frac{r^4}{\bar{\Lambda}^4}\,\left[\frac{1}{\gamma^2}\,\frac{1}{\bar{\Lambda}^4}-\frac{1}{r^2}\,\left(1-\alpha-\frac{2\,M}{r}+\frac{r^2}{\ell^2_p}\right)\right].\label{dd4}
\end{equation}

Transforming to a new variable via $r=\frac{1}{u}$ into the Eq. (\ref{dd4}), we find the photon trajectory equation as follows:
\begin{equation}
    \left(\frac{du}{d\phi}\right)^2=\frac{1}{\gamma^2}\,\frac{1}{\bar{\Lambda}^8}-\frac{u^2}{\bar{\Lambda}^4}\,\left(1-\alpha-2\,M\,u+\frac{1}{u^2\,\ell^2_p}\right),\quad\quad \bar{\Lambda}=1+\frac{B^2_0}{u^2}.\label{dd5}
\end{equation}

From the equation above (\ref{dd5}), it is clear that several factors influence the photon trajectories when orbiting in the vicinity of a magnetized BH. These factors include the magnetic field strength $B_0$, the radius of curvature $\ell_p$, and the CS parameter $\alpha$. Additionally, the mass of the BH $M$ also affects the photon trajectory in the given gravitational field. In the limit where $B_0 = 0$, meaning that there is no magnetic field influence on the photon particles, the photon trajectories reduce to those corresponding to the Letelier-AdS BH solution. 

Now, we determine the force on the photon particles and analyze how various parameters influence the motion of photon particles as they move under the influence of the gravitational field generated by the magnetized BH. This force is expressible in terms of the effective potential as:
\begin{equation}
    \mathrm{F}_\text{ph}=-\frac{1}{2}\,\frac{\partial V_\text{eff}}{\partial r},\label{force}
\end{equation}                          
where $V_\text{eff}$ is the effective potential for null geodesics and is given in Eq. (\ref{dd3cc}). 

Thereby, using the expression (\ref{dd3cc}) into the Eq. (\ref{force}), we find
\begin{eqnarray}
    \mathrm{F}_\text{ph}=\frac{(1+B^2_0\,r^2)^3\,\mathrm{L}^2}{r^3}\,\left[1-\alpha-\frac{3\,M}{r}-3\,(1-\alpha)\,B^2_0\,r^2+5\,M\,B^2_0\,r-\frac{4\,B^2_0}{\ell^2_p}\,r^4\right].\label{dd6}
\end{eqnarray}

From the expression in Eq.  (\ref{dd6}), it is evident that various factors, such as the magnetic field of strength $B_0$, radius of the curvature $\ell_p$, and the CS parameter $\alpha$ influences the motion of photon particles when moving in the given gravitational field. Additionally, the mass of the BH $M$ also influence the photon dynamics. In the limit where $B_0=0$,  meaning there is no magnetic field influence on the motion of photon particles, the force decreases and the result reduce to those corresponding to the Letelier-AdS BH solution.

\begin{figure}[ht!]
    \centering
    \subfloat[$B_0=0.1$]{\centering{}\includegraphics[width=0.45\linewidth]{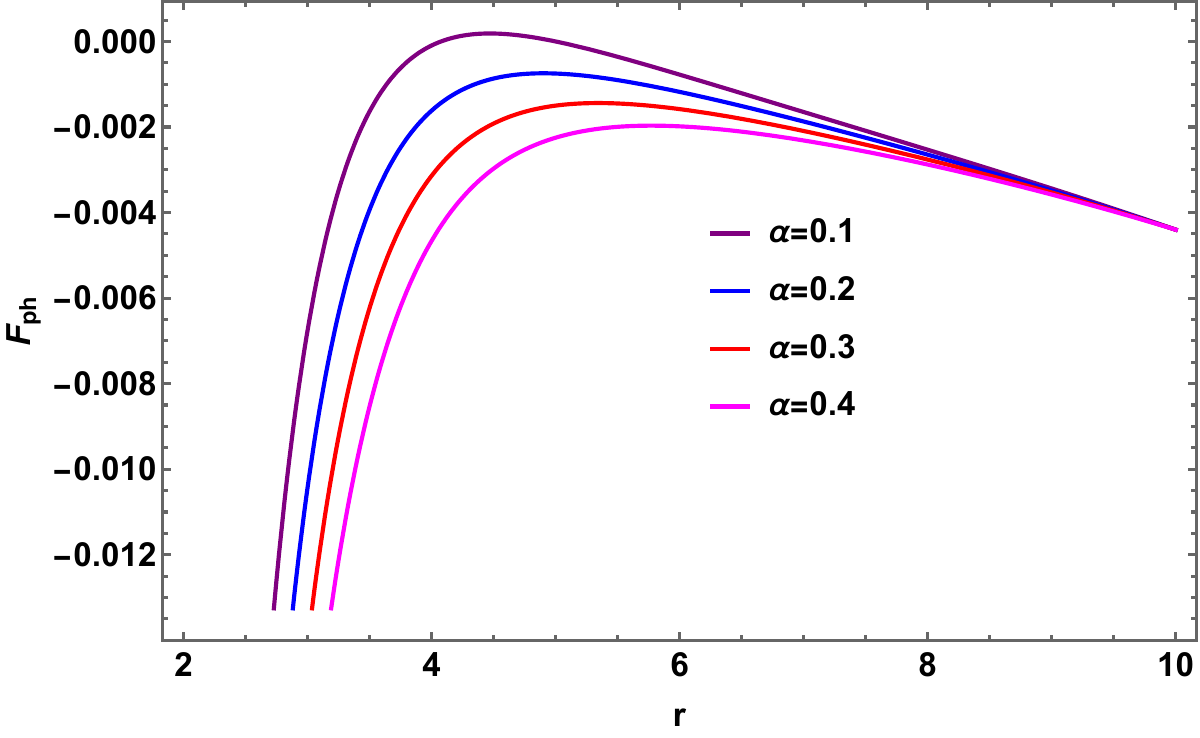}}\quad\quad
    \subfloat[$\alpha=0.1$]{\centering{}\includegraphics[width=0.45\linewidth]{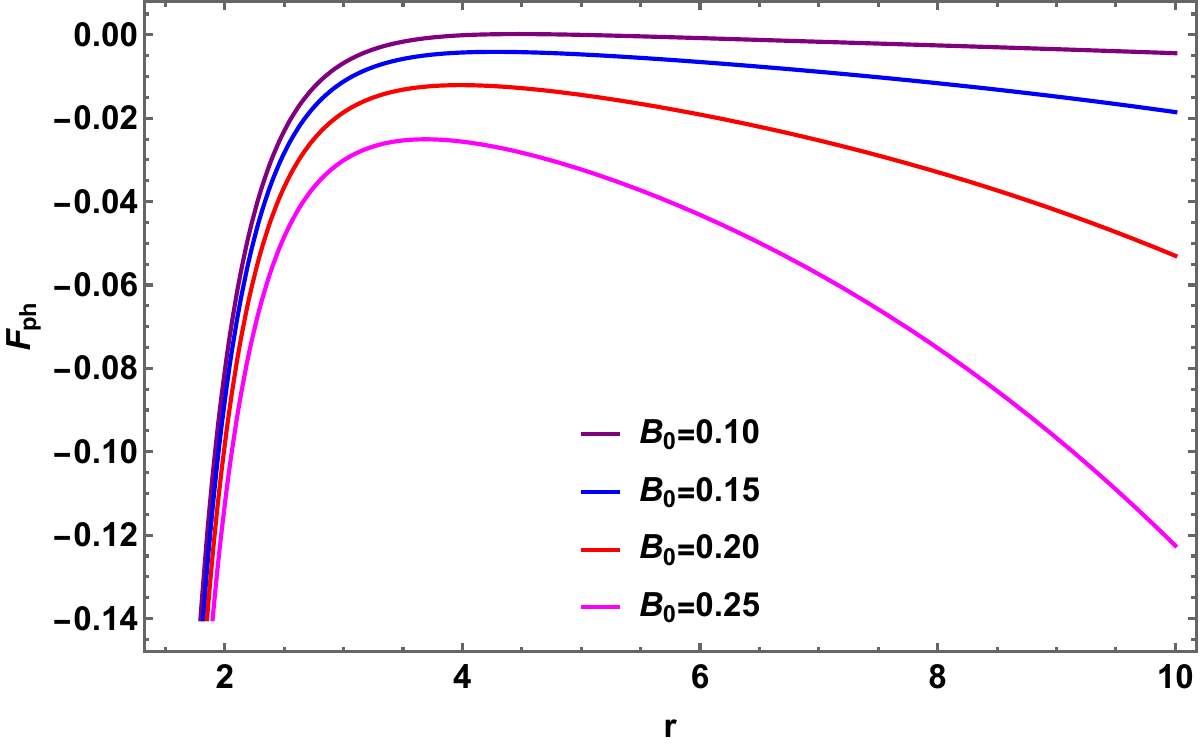}}\\
    \subfloat[]{\centering{}\includegraphics[width=0.45\linewidth]{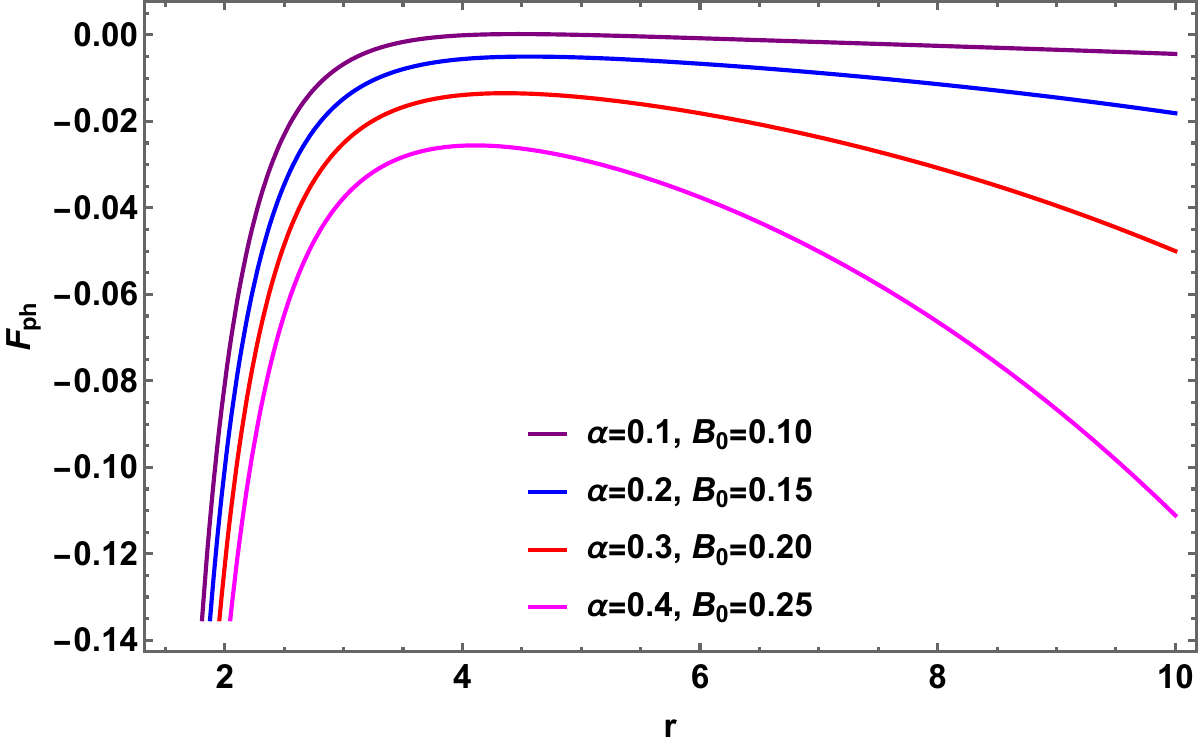}} 
    \caption{The behavior of force on photon particles by varying $\alpha$ and $B_0$. Here $M=1$, $\mathrm{L}=1$, and $\ell_p=10$.}
    \label{fig:force1}
\end{figure}

\begin{figure}[ht!]
    \centering
    \includegraphics[width=0.5\linewidth]{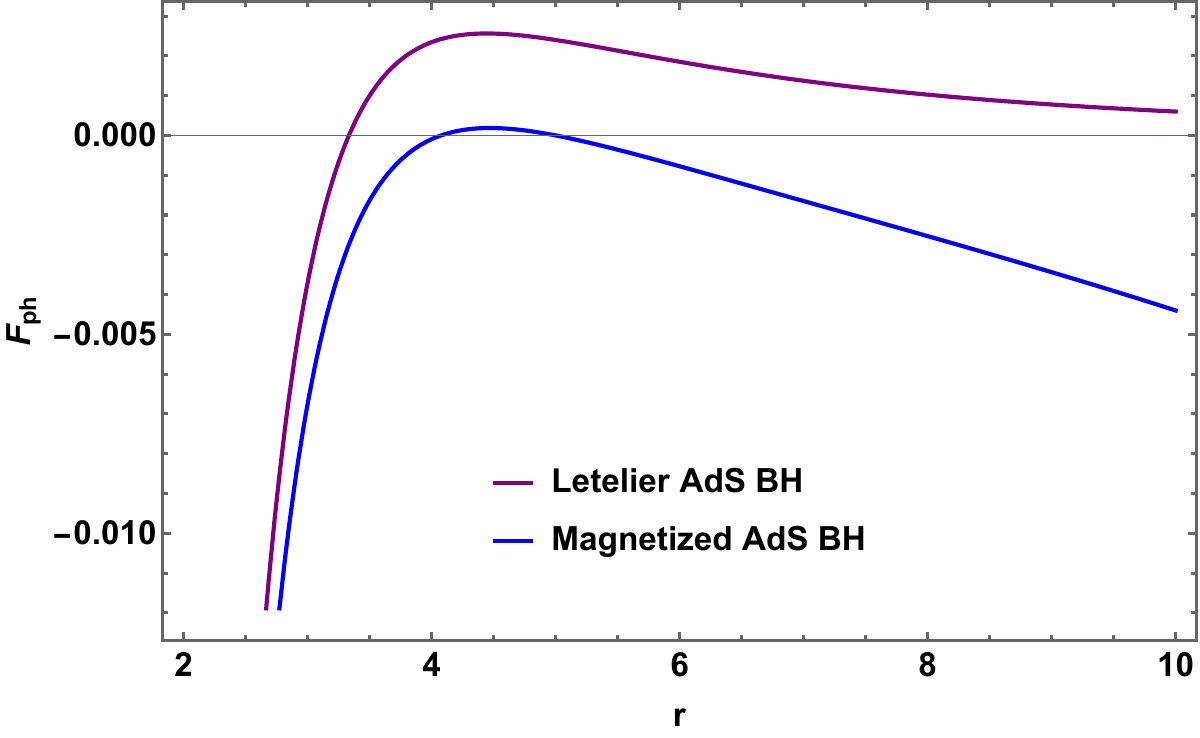}
    \caption{A comparison of the effective potential for charge particle. Here $M=0.8$, $\mathrm{L}=1$, and $\ell_p=10$. Purple color: $\alpha=0.1, B_0=0$, blue color: $B_0=0.1, \alpha=0.1$.}
    \label{fig:force2}
\end{figure}

In Fig. \ref{fig:force1}, we present the behavior of the force acting on photon particles by the gravitational fieldgenerated by the magnetized BH under variations of the cosmic string (CS) parameter $\alpha$ and the magnetic field strength $B_0$. In panels (a) to (c), we observe that increasing the CS parameter from $\alpha = 0.1$, the magnetic field strength $B_0$, or their combination ($\alpha$, $B_0$), results in a decrease in the force experienced by photon particles as a function of the radial coordinate $r$. This behavior suggests that small variations in $\alpha$ and $B_0$, or their combined effect, reduce the strength of the gravitational field generated by the magnetized BH, thereby diminishing its influence on photon dynamics.

Similarly, in Figure \ref{fig:force2}, we present a comparison of the force acting on photon particles with and without the influence of a magnetic field. We observe that a slight increase in the magnetic field strength, from $B_0 = 0$ to $B_0 = 0.1$, results in a more significant variation in the force experienced by the photon particles. In these Figures \ref{fig:force1}--\ref{fig:force2}, the BH mass is fixed at $M = 1$, the angular momentum $\mathrm{L}=1$, and the AdS radius at $\ell_p = 10$. These parameters collectively govern the motion of photon particles and influence whether they are captured by or escape from the gravitational field of the BH.

The period of circular orbits encompasses the time required for a particle to complete one full revolution around the circular path. The formulas for calculating the period of a circular orbit in proper ($T_{\tau}$) and coordinate times ($T_t$) are derived in Ref. \cite{SF}. In our case, following the similar approach, we find these times are
\begin{eqnarray}
    T_{\tau}=\frac{2\,\pi\,r^2_c}{\mathrm{L}},\label{dd7}
\end{eqnarray}
where $\mathrm{L}$ is the conserved angular momentum. And
\begin{equation}
    T_t=2\,\pi\,|\beta_c|=\frac{2\,\pi\,r_c}{(1+B^2_0\,r^2_c)^2\,\sqrt{1-\alpha-\frac{2\,M}{r_c}+\frac{r^2_c}{\ell^2_p}}},\label{dd8}
\end{equation}
where $r_c$ is nothing bu the photon sphere radius that can be determined from Eq. (\ref{dd3ee}).

Finally, we focus on the stability of circular orbits. The stability of these orbits can be determined using a physical quantity known as the Lyapunov exponent. When the Lyapunov exponent is imaginary, the system exhibits oscillatory behavior, indicating the presence of stable circular orbits. This implies that small perturbations in the photon trajectory lead to periodic oscillations around the orbit, ensuring the orbit remains stable. On the other hand, if the Lyapunov exponent is real and positive, the system is characterized by chaotic motion. In this case, nearby photon trajectories diverge exponentially, signaling instability in the orbits. 

The Lyapunov exponent in terms of the effective potential for circular orbit is defined by \cite{VC}
\begin{equation}
    \lambda^\text{null}_L=\sqrt{-\frac{V''_\text{eff}(r_c)}{2\,\dot{t}^2}},\label{cond1}
\end{equation}
where $\dot{t}$ is given in Eq. (\ref{dd2}).

Substituting the effective potential (\ref{dd1}) and after simplification, we find
\begin{eqnarray}
    \lambda^\text{null}_L&=&(1+B^2_0\,r^2_c)^2\,\sqrt{1-\alpha-\frac{2\,M}{r_c}+\frac{r^2_c}{\ell^2_p}}\times\nonumber\\
    &&\sqrt{\frac{1}{r^2_c}\,\left(1-\alpha+\frac{2\,M}{r_c}\right)-\frac{4\,B^2_0\,(B^2_0\,r^3_c+7\,B^2_0\,r^2_c+r_c+1)}{(1+B^2_0\,r^2_c)^2}\,\left(1-\alpha-\frac{2\,M}{r_c}+\frac{r^2_c}{\ell^2_p}\right)}.\label{cond2}
\end{eqnarray}
where we have used the circular orbit condition $V_\text{eff}'(r=r_c)=0$ that simplifies to
\begin{equation}
    r_c\,\mathcal{F}'(r_c)=2\,\mathcal{F}(r_c)-4\,r_c\,\mathcal{F}(r_c)\,\frac{(\tilde{\Lambda})'}{\tilde{\Lambda}},\label{cond3}
\end{equation}
where prime denotes partial derivative w. r. t. $r$. 

In the limit where $B_0=0$, that is, absence of the magnetic field, we find the Lyapunov exponent
\begin{eqnarray}
    \lambda^\text{null}_L&=&\frac{1}{r_c}\,\sqrt{\left(1-\alpha-\frac{2\,M}{r_c}+\frac{r^2_c}{\ell^2_p}\right)\,\left(1-\alpha+\frac{2\,M}{r_c}\right)}.\label{cond4}
\end{eqnarray}

Equation (\ref{cond3}) represents the Lyapunov exponent for circular null geodesics in the background of the magnetized AdS BH metric, whereas Equation (\ref{cond4}) corresponds to the Lyapunov exponent for circular null geodesics in the background of the Letelier AdS BH metric. From expressions (\ref{cond3})--(\ref{cond4}), we observe that the presence of magnetic field causes a shift in the Lyapunov exponent in the current study compared to the Letelier BH metric.

\begin{figure}[ht!]
    \centering
    \subfloat[$B_0=0.1$]{\centering{}\includegraphics[width=0.45\linewidth]{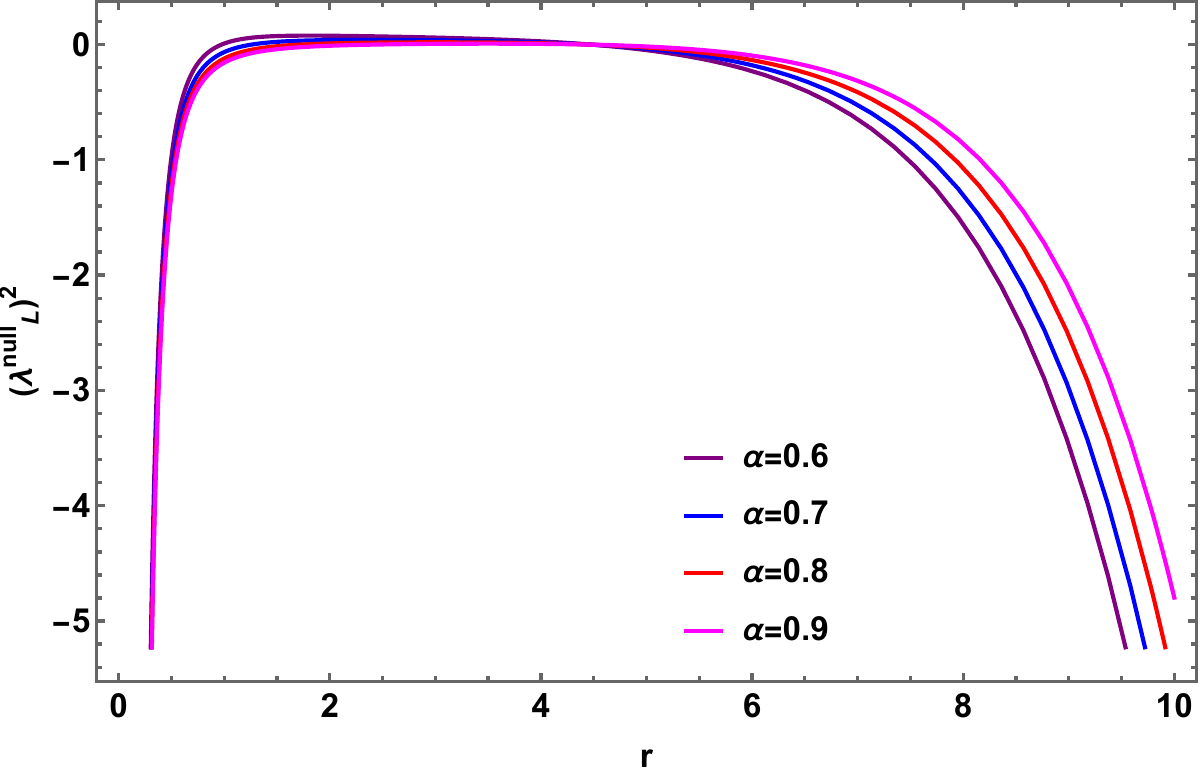}}\quad\quad
    \subfloat[$\alpha=0.1$]{\centering{}\includegraphics[width=0.45\linewidth]{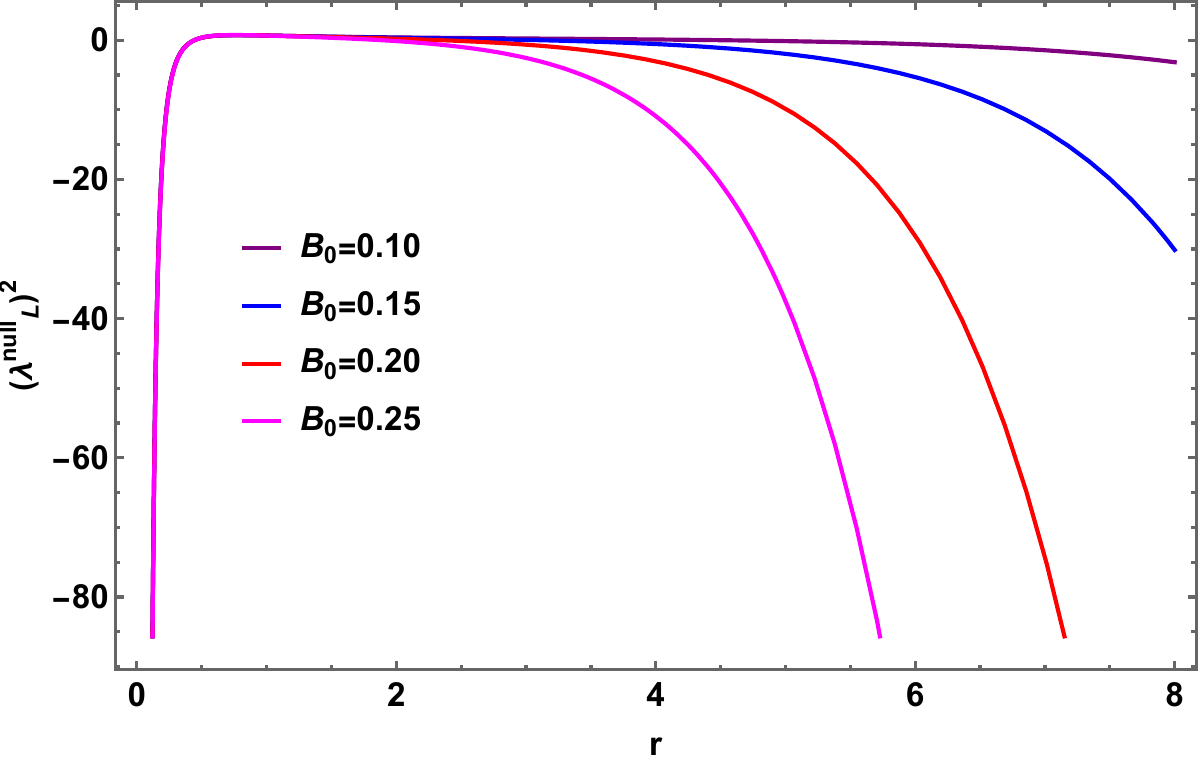}}\\
    \subfloat[]{\centering{}\includegraphics[width=0.45\linewidth]{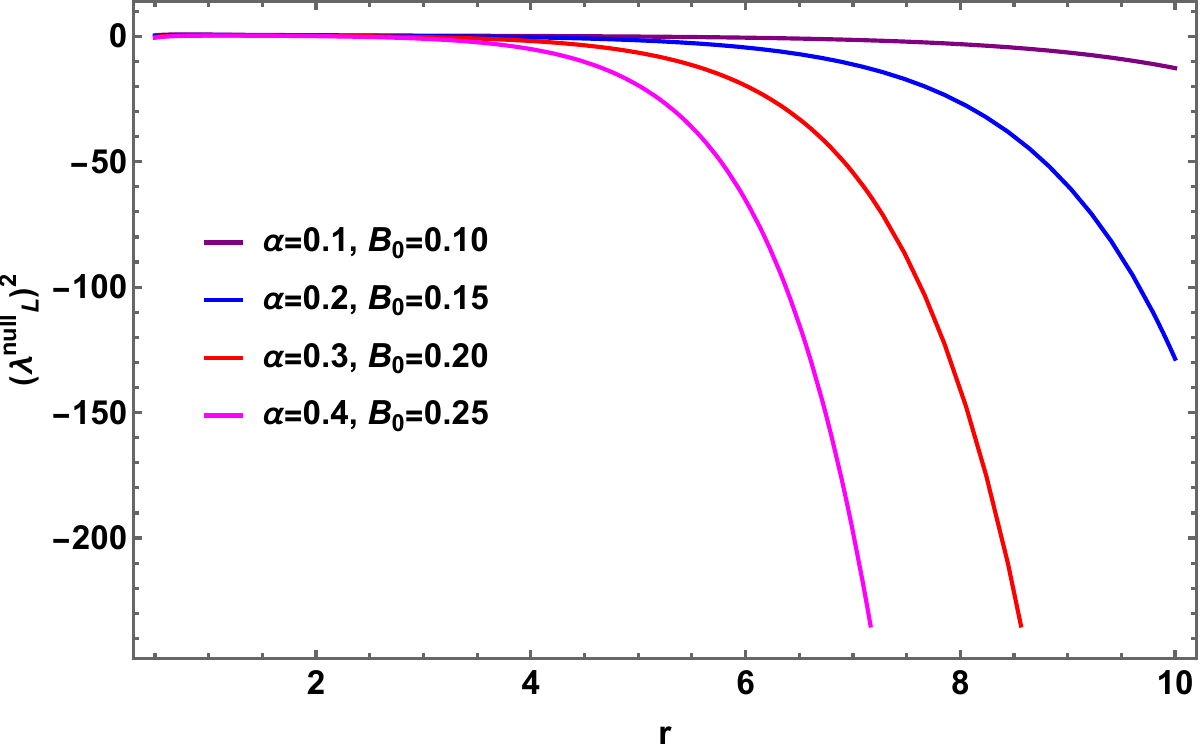}} 
    \caption{The square of the Lyapunov exponent $\lambda^\text{null}_L$ by varying $\alpha$ and $B_0$ as a function of $r=r_c \neq 0$. Here $M=0.2$, and $\ell_p=10$.}
    \label{fig:lyapunov1}
\end{figure}

\begin{figure}[ht!]
    \centering
    \includegraphics[width=0.45\linewidth]{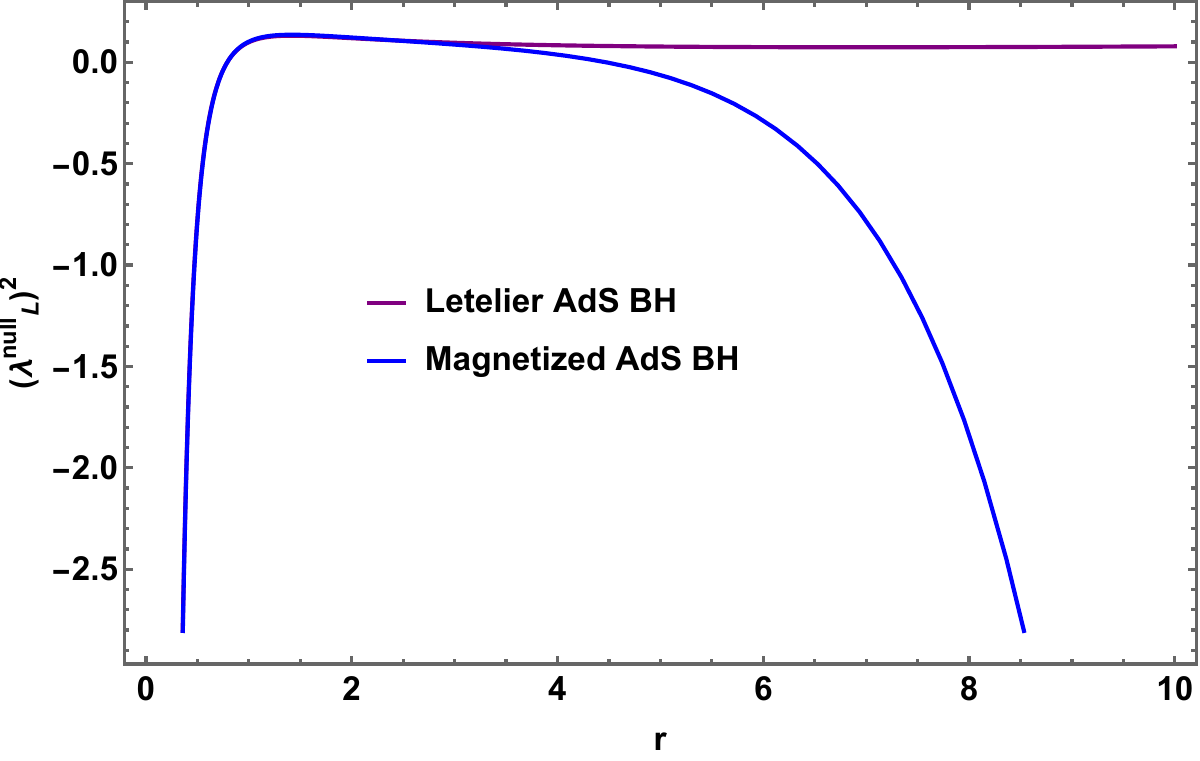}
    \caption{A comparison of the square of the Lyapunov exponent $\lambda^\text{null}_L$. Here $M=0.2$, and $\ell_p=10$. Purple color: $\alpha=0.5, B_0=0$, blue color: $B_0=0.1, \alpha=0.5$.}
    \label{fig:lyapunov2}
\end{figure}

In Fig. \ref{fig:lyapunov1}, we present the square of the Lyapunov exponent for circular null orbits under variations of the cosmic string (CS) parameter $\alpha$ and the magnetic field strength $B_0$. We observe that increasing the CS parameter from $\alpha = 0.6$, the magnetic field strength $B_0$, or their combination ($\alpha$, $B_0$), results in negative values of the squared Lyapunov exponent as a function of the circular orbit radius $r = r_c \neq 0$. This behavior indicates that higher values of $\alpha$ and small rise in $B_0$, or their combined effect, lead to an imaginary Lyapunov exponent, which corresponds to stable circular null orbits.

Similarly, in Fig. \ref{fig:lyapunov2}, we present a comparison of the square of the Lyapunov exponent for circular null orbits with and without the influence of a magnetic field. We observe that a slight increase in the magnetic field strength, from $B_0 = 0$ to $B_0 = 0.1$, leads to negative values of the squared Lyapunov exponent as a function of the circular orbit radius $r = r_c \neq 0$. This suggests that the circular null geodesics in the present study are more stable compared to those in the standard AdS BH space-time with CS effects. In these figure, the BH mass is fixed at $M = 0.2$, and the AdS radius at $\ell_p = 10$.

\section{Magnetized Letelier BH Shadow} \label{isec6}

\begin{figure}[ht!]
    \centering
    \includegraphics[width=0.45\linewidth]{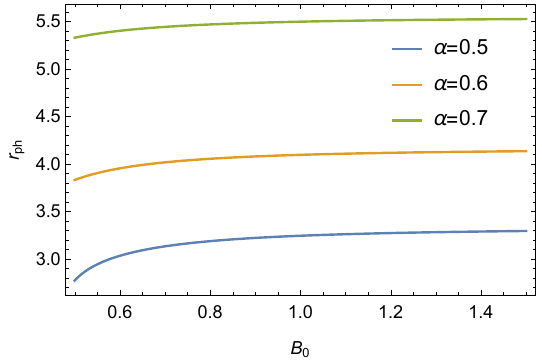}
    \caption{Variation  of  the  photon sphere
$r_{ph}$ of the magnetized AdS Letelier BH with the $B_0$ parameter for various CS parameter $\alpha$; here, $M=1$ and $\ell_p=100$. }
    \label{figa77}
\end{figure}

To calculate the shadow radius, one must first determine the photon sphere radius ($r_{ph}$). To this end, we introduce the effective potential for null geodesics in an equatorial plane: \begin{equation}
    V(r)= \Lambda^4\,\frac{ \mathcal{F}(r)}{r^2}
\end{equation}
A photon sphere radius $r_{ph}$ around the BH is given by the equation (\ref{dd3ee})  $V'(r=r_{ph})=0,$ 
The complexity of Eq. (\ref{dd3ee}) represents the delicate interplay between the Letelier parameter $\alpha$ and the strength of the external magnetic field parameters $B_0$ in defining the photon sphere position. \\  To solve  Eq. (\ref{dd3ee}), we use a numerical technique to get the photon orbit radii $r_{ph}$.  Table \ref{taba33} presents these numerical results, revealing the influence of the parameters ($\alpha, B_0$) on photon sphere radii.
\begin{center}
\begin{tabular}{|c|c|c|c|c|c|c|}
 \hline 
 \multicolumn{7}{|c|}{ $r_{ph}$ }
\\ \hline 
& $B_{0}=0.5$ & $0.6$ & $0.7$ & $0.8$ & $0.9$ & $1$ \\ \hline
$\alpha =0.5$ & $2.76147$ & $3.02489$ & $3.12491$ & $3.17944$ & $3.21343$ & $%
3.23634$ \\ 
$0.6$ & $3.80906$ & $3.93285$ & $3.99613$ & $4.03384$ & $4.0584$ & $4.07537$
\\ 
$0.7$ & $5.26271$ & $5.33638$ & $5.37771$ & $5.40344$ & $5.42063$ & $5.43271$ \\ 
 
 \hline
\end{tabular}
\captionof{table}{Numerical results for the photon sphere  with various values of the CS $\alpha $ and the  magnetic field parameter $B_0$. Here $M=1$ and $\ell_p=300$.} \label{taba33}
\end{center}
Our numerical analysis in Table \ref{taba33} demonstrates that  as the magnetic field parameter $B_0$ increases, the photon sphere radius $r_{ph}$ exhibits a systematic increase. Similar trend is observed for the CS parameter $\alpha$ (see Fig. \ref{figa77}).

\begin{figure}[ht!]
    \centering
    \includegraphics[width=0.45\linewidth]{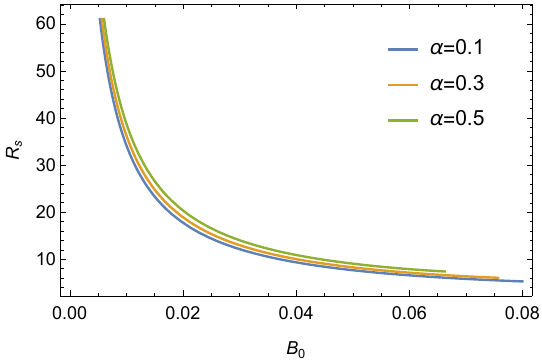}
    \caption{Variation  of  the  shadow  observable
$R_s$ of the magnetized AdS Letelier BH with the $B_0$ parameter for various CS parameter $\alpha$; here, $M=1$ and $\ell_p=100$. }
    \label{figa7}
\end{figure}

After getting the photon orbit radii, we may calculate the shadow radii directly using 
 \begin{equation}
    R_s\rightarrow \beta_c=\frac{1}{\sqrt{V(r=r_{ph})}}. \label{shad1}
\end{equation}
Equation (\ref{shad1}) uses the shadow radius formula to calculate the apparent size of the BH from a distance in the equatorial plane. Physically, it relates to a critical impact parameter that distinguishes capture and scattering orbits for light rays approaching the BH. For viewers at infinity, this essential impact parameter appears as the radius of a black circular patch on a bright background—the BH shadow. \\ To have a better understanding of the impact of the external magnetic field, we tabulate values for the  shadow radius in Table  \ref{taba3}. 
\begin{center}
\begin{tabular}{|c|c|c|c|c|c|c|}
 \hline 
 \multicolumn{7}{|c|}{ $R_{s}$ }
\\ \hline 
& $B_{0}=0.01$ & $0.02$ & $0.03$ & $0.04$ & $0.05$ & $0.06$ \\ \hline
$\alpha =0.1$ & $34.2276$ & $17.7339$ & $12.1227$ & $9.32031$ & $7.65394$ & $%
6.56198$ \\ 
$0.2$ & $36.3091$ & $18.8995$ & $12.9668$ & $10.0091$ & $8.25859$ & $7.12293$
\\ 
$0.3$ & $38.8253$ & $20.331$ & $14.0177$ & $10.8809$ & $9.04094$ & $7.87539$ \\ 
 
 \hline
\end{tabular}
\captionof{table}{Numerical results for the shadow radius with various values of the CS $\alpha $ and the  magnetic field parameter $B_0$. Here $M=1$ and $\ell_p=300$.} \label{taba3}
\end{center}
It is clearly shown in Table \ref{taba3} that the shadow radius decreases with increasing magnetic field parameter, but increases with CS parameter. Furthermore, we plot the shadow radius viruses the magnetic field parameter for different values of CS parameter  $\alpha$ (see Fig. \ref{figa7}). The figure indicates that when $\alpha$ increases, so does the shadow radius whereas it decreases with $B_0$. This suggests that the external magnetic field decreases the shadow radius.

To represent the actual shadow of the magnetized Letelier BH as seen from an observer's perspective, we introduce celestial coordinates, $X$ and $Y$
\begin{equation}
X=\lim_{r_{\mathrm{o}}\rightarrow \infty }\left( -r_{\mathrm{o}}^{2}\sin
\theta _{\mathrm{o}}\frac{d\varphi }{dr}\right) ,
\end{equation}%
\begin{equation}
Y=\lim_{r_{\mathrm{o}}\rightarrow \infty }\left( r_{\mathrm{o}}^{2}\frac{%
d\theta }{dr}\right) .
\end{equation}

For a static observer at large distance, i.e. at  $r_{\mathrm{o}}\rightarrow \infty $ in the equatorial plane $\theta {\mathrm{o}}=\pi /2$, the celestial coordinates simplify to 
\begin{equation}
X^{2}+Y^{2}=R_s^{2}.
\end{equation}

\begin{figure}[ht!]
    \centering
    \includegraphics[scale=0.6]{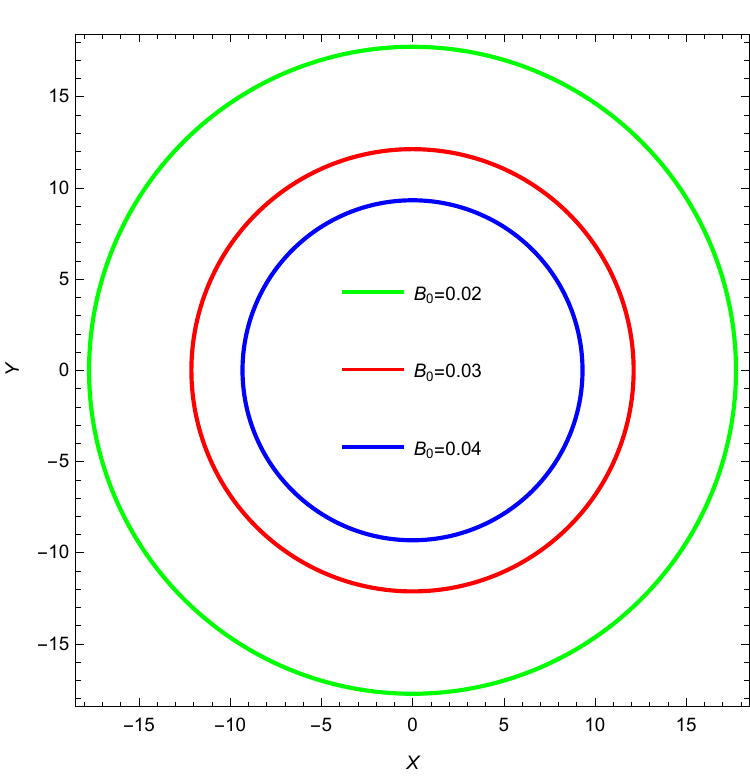} \includegraphics[scale=0.6]{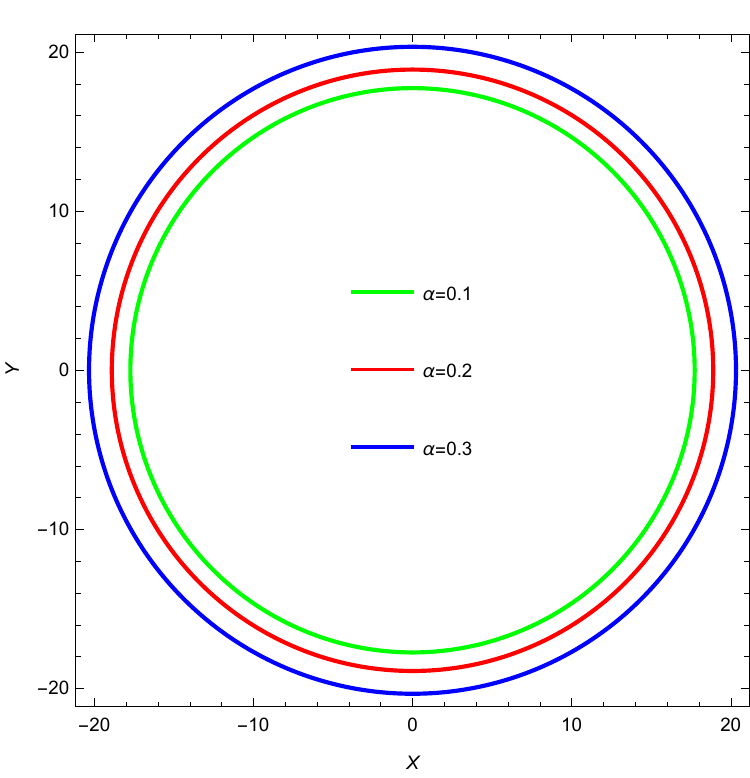}
    \caption{BH shadows  for different values of $B_0$  keeping  $\alpha = 0.1$ (left) and for different values of $\alpha$ keeping $B_0=0.02$ (right).}
    \label{ps25}
\end{figure}

Now we'll show how the magnetic field and CS affect the BH shadows. Fig. \ref{ps25} depicts the impact of a magnetic field and shows that for smaller values of $B_0$ parameters, the shadows have bigger size. Shadows, on the other hand, have smaller radii as $\alpha$ decreases.

\section{Conclusions} \label{isec7}

{\color{black}In this study, we conducted a comprehensive investigation of the magnetized Letelier BH in AdS spacetime, focusing on the motion of charged, neutral, and light-like particles, as well as the BH shadow. Our analysis revealed how the key parameters-namely, the magnetic field strength $B_0$, the CS parameter $\alpha$, and the AdS radius $\ell_p$-significantly affect the dynamics of particles and the optical properties of this spacetime.

We began our investigation by deriving the metric for the magnetized Letelier BH in AdS spacetime, as given in Eq. (\ref{bb1}), and establishing its relationship to other well-known spacetime solutions. The magnetic field components, measured by ZAMOs and expressed in Eqs. (\ref{b1}) and (\ref{b2}), provided insights into the electromagnetic environment surrounding the BH. Our analysis of the metric function $\mathcal{F}(r)$, illustrated in Fig. \ref{figa1}, demonstrated that the CS parameter $\alpha$ plays a crucial role in determining the BH horizon, while the magnetic field strength $B_0$ does not affect the horizon position. This finding underscores the distinct influences of these parameters on the spacetime geometry.

For charged particles moving in the equatorial plane, we derived the effective potential given by Eq. (\ref{cc5}), which incorporates the combined effects of the magnetic field, CS, and AdS background. Our analysis, visualized in Fig. \ref{fig:potential1}, demonstrated that increasing the CS parameter $\alpha$ leads to a decrease in the effective potential, indicating a weakening of the gravitational influence. In contrast, increasing the magnetic field strength $B_0$ results in an enhancement of the effective potential, suggesting a strengthening of the gravitational interaction. The comparison presented in Fig. \ref{fig:potential2} highlighted the significant modifications introduced by the magnetic field to the effective potential experienced by charged particles.

Our investigation of particle trajectories, illustrated in Fig. \ref{Fig:traj}, revealed the complex interplay between the gravitational and magnetic influences. We observed that the presence of a non-zero magnetic field parameter ($B_0 \neq 0$) introduces an attractive component that can destabilize particle orbits, transforming bound orbits into captured orbits. Furthermore, the combined influence of the magnetic field and CS parameter can shift particle orbits from bound to captured states, demonstrating the profound impact of these parameters on particle dynamics. For neutral particles, we derived expressions for the specific angular momentum and energy in circular orbits, given by Eqs. (\ref{ss12}) and (\ref{ss13}), respectively. The plots in Fig. \ref{fig:neutral-particle} illustrated that both the specific angular momentum and energy increase with higher values of the CS parameter $\alpha$ and the magnetic field strength $B_0$. This finding indicates that these parameters increase the energy requirements for particles to maintain stable circular orbits around the BH.

Our analysis of null geodesics provided insights into the optical properties of the magnetized Letelier BH in AdS spacetime. The effective potential for photon particles, given by Eq. (\ref{dd3bb}), exhibited similar dependencies on the CS parameter and magnetic field strength as observed for material particles. Fig. \ref{fig:potential3} showed that increasing the CS parameter weakens the gravitational influence on photon particles, while increasing the magnetic field strength enhances it. The comparative analysis in Fig. \ref{fig:potential4} emphasized the substantial modifications introduced by the magnetic field to the effective potential experienced by photons. We determined the photon sphere radius by solving Eq. (\ref{dd3ee}) numerically for various parameter combinations. Table \ref{taba33} and Fig. \ref{figa77} demonstrated that the photon sphere radius increases with both the magnetic field parameter $B_0$ and the CS parameter $\alpha$. This result indicates that these parameters push the region of unstable circular photon orbits further away from the BH, affecting the optical appearance of the BH to distant observers.

The force acting on photon particles, derived in Eq. (\ref{dd6}), was analyzed for different parameter values in Fig. \ref{fig:force1}. We found that increasing either the CS parameter or the magnetic field strength leads to a decrease in the force experienced by photon particles, suggesting a reduction in the strength of the gravitational field's influence on photon dynamics. The comparison presented in Fig. \ref{fig:force2} highlighted the significant variations introduced by the magnetic field to the force acting on photons. For the stability of circular null geodesics, we calculated the Lyapunov exponent as given by Eq. (\ref{cond2}). The analysis presented in Fig. \ref{fig:lyapunov1} showed that increasing either the CS parameter or the magnetic field strength results in negative values of the squared Lyapunov exponent, indicating stable circular null orbits. This finding, also supported by the comparison in Fig. \ref{fig:lyapunov2}, suggests that the presence of a magnetic field stabilizes circular photon orbits compared to the standard AdS BH spacetime with CS effects.

Finally, we examined the BH shadow, calculating the shadow radius using Eq. (\ref{shad1}). Table \ref{taba3} and Fig. \ref{figa7} revealed that the shadow radius decreases with increasing magnetic field parameter but increases with the CS parameter. This result indicates that the external magnetic field reduces the apparent size of the BH shadow, while the CS effect enlarges it. The visualization of the BH shadows for different parameter values in Fig. \ref{ps25} provided a clear illustration of how these parameters affect the optical appearance of the BH to distant observers. Throughout our analysis, we observed consistent patterns in how the CS parameter and magnetic field strength influence various physical quantities. The CS parameter generally weakens the gravitational interaction, resulting in larger horizon and photon sphere radii, decreased effective potentials, and increased shadow sizes. In contrast, the magnetic field strength introduces additional attractive components that can destabilize particle orbits while simultaneously stabilizing circular photon orbits. These findings highlight the complex and sometimes counterintuitive effects that arise from the combination of CS, magnetic fields, and AdS background in BH spacetimes. Overall, the magnetized Letelier BH in AdS spacetime represents a more realistic model for astrophysical BHs, which often exist in environments with magnetic fields and potentially exotic matter distributions such as cosmic strings. The distinct effects of these parameters on particle dynamics and optical properties offer potential detectable discriminators that could be used to test different BH models against astronomical data \cite{izz29,izz30}.

Looking forward, several promising directions for future research emerge from our study. First, extending our analysis to include rotating BHs would provide a more complete picture of astrophysical BHs, which typically possess significant angular momentum \cite{izz31}. Second, investigating the thermodynamic properties of magnetized Letelier BHs in AdS spacetime could reveal interesting connections to the AdS/CFT correspondence and quantum gravity theories \cite{izz32,izz32x}. Third, simulating the accretion processes and jet formation around these BHs would enable direct comparisons with observational data from facilities like the EHT and future X-ray observatories \cite{EventHorizonTelescope:2024uoo}. Finally, exploring the quasinormal modes and gravitational wave signatures of these BHs could provide additional observational probes of the spacetime geometry, especially in the era of multi-messenger astronomy \cite{izz32xx}. }

{\small

\section*{Acknowledgments}

 F.A. gratefully acknowledges the Inter University Centre for Astronomy and Astrophysics (IUCAA), Pune, India, for the conferment of a visiting associateship.  \.{I}.~S. extends sincere thanks to TÜBİTAK, ANKOS, and SCOAP3 for their support in facilitating networking activities under COST Actions CA22113, CA21106, and CA23130. S.S is supported by the National Natural Science Foundation of China under Grant No. W2433018.

\section*{Data Availability Statement}

There are no new data associated with this article.
}

\end{document}